\documentclass[a4paper,fleqn,usenatbib]{mnras}

\usepackage[T1]{fontenc}
\usepackage{ae,aecompl}

\usepackage{graphicx} 
\usepackage{amsmath}  
\usepackage{amssymb}  

\title[]{Galactic Simulations of r-process Elemental Abundances}

\author[C. J. Haynes et al.]{
Christopher J. Haynes,$^{1}$\thanks{E-mail: c.haynes@herts.ac.uk}
Chiaki Kobayashi,$^{1}$
\\
$^{1}$Centre for Astrophysics Research, University of Hertfordshire, College Lane, Hatfield, AL10 9AB, UK\\
}

\date{Accepted XXX. Received YYY; in original form ZZZ}

\pubyear{2018}

\begin{document}
\label{firstpage}
\pagerange{\pageref{firstpage}--\pageref{lastpage}}
\maketitle

\begin{abstract}
We present the distributions of elemental abundance ratios using chemodynamical simulations which include four different neutron capture processes: magneto-rotational supernovae, neutron star mergers, neutrino driven winds, and electron capture supernovae. We examine both simple isolated dwarf disc galaxies and cosmological zoom-in simulations of Milky Way-type galaxies, and compare the [Eu/Fe] and [Eu/$\mathrm{\alpha}$] evolution with recent observations, including the HERMES-GALAH survey. We find that neither electron-capture supernovae or neutrino-driven winds are able to adequately produce heavy neutron-capture elements such as Eu in quantities to match observations. Both neutron-star mergers and magneto-rotational supernovae are able to produce these elements in sufficient quantities. Additionally, we find that the scatter in [Eu/Fe] and [Eu/$\mathrm{\alpha}$] at low metallicity ([Fe/H] < $-1$) and the [Eu/(Fe, $\mathrm{\alpha}$)] against [Fe/H] gradient of the data at high metallicity ([Fe/H] > $-1$) are both potential indicators of the dominant r-process site. Using the distribution in [Eu/(Fe, $\mathrm{\alpha}$)] -- [Fe/H] we predict that neutron star mergers alone are unable to explain the observed Eu abundances, but may be able to together with magneto-rotational supernovae. 
\end{abstract}




\section{Introduction} 

Elements heavier than iron are produced by neutron-capture processes. There are two processes, slow and rapid (s-process and r-process), which differ in both site and nucleosynthesis yields but share the same basic mechanism: seed nuclei absorb neutrons which subsequently decay into protons via beta minus decay to increase the atomic number. For the s-process to occur the timescale for neutron absorption must be much longer than the timescale for beta minus decay. The steady neutron flux required for this is thought to result from He-burning, both in the centres of massive stars (producing the weak s-process elements up to roughly A<90, see \citealt{PrantzosEtAl90}, \citealt{Frischknecht16}) and in the He-burning shells of low mass AGB stars (comprising the main s-process elements at A>90, see \citealt{SmithLambert90} and \citealt{Kaeppeler11}). 

In this paper we focus instead on the r-process, where a short neutron capture timescale allows for multiple neutron capture events to occur prior to beta minus decay. This particular method of nucleosynthesis is thought to be responsible for the assembly of the heaviest and most neutron-rich elements. The r-process requires extremely high neutron densities to operate and despite the limit this places on potential astrophysical sites the origin of the process remains a key unanswered question in galactic archaeology. Historically, there have been four theoretical sites: the dynamic ejecta of neutron star (NS) mergers \citep[e.g.,][]{Lattimer74,FreiburghausEtAl99}, magneto-rotational supernovae \citep[MRSNe, e.g.,][]{Cameron03}, Electron capture supernovae \citep[ECSNe, e.g.,][]{Wanajo11}, and neutrino driven winds \citep[NUW, e.g.,][]{Wanajo01}. 

\begin{figure*} 
\centering 
\includegraphics[scale=0.42]{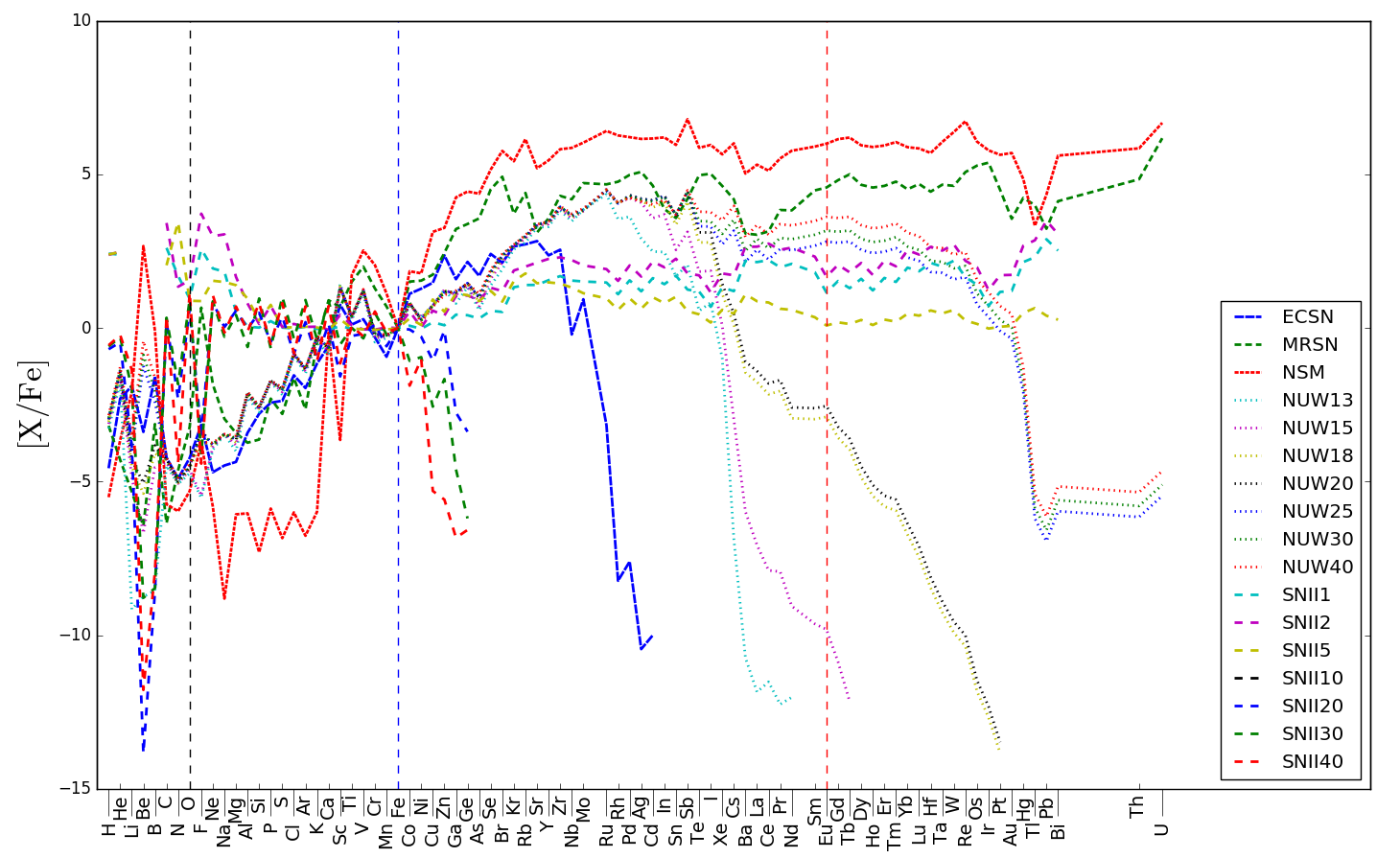} 
\caption{The nucleosynthesis yields relative to iron that we use in our simulations. SNII and AGB contribution for stars of 1 to 40 $\mathrm{M_{\odot}}$ (SNII1 - SNII40) are shown as dashed lines. The yields for neutrino driven winds from stars of 13 to 40 $\mathrm{M_{\odot}}$ are shown as dotted lines (NUW13 - NUW40). ECSNe, MRSNe and NS mergers are shown with the blue, green and red dashed lines respectively. Vertical dashed lines are shown at oxygen, iron and europium for ease of comparison.} 
\label{NucYields} 
\end{figure*} 

Because these events are intrinsically rare, it is necessary to consider inhomogeneous enrichment in modelling of the evolution of r-process elements. In previous works, stochastic chemical evolutions have been used (\citealt{Cescutti15}, \citealt{Wehmeyer15}, \citealt{Ishimaru17}). The disadvantage of these models is that the star formation history has to be assumed for predicting the elemental abundances. In this paper, we use chemodynamical, hydrodynamical simulations, including relevant baryon physical processes (such as star formation and supernova feedback), to simulate the formation and evolution of Milky Way-type galaxies from cosmological initial conditions. These processes can play an important role in helping to determine which, if any, of these sites are dominant as they allow us to compare theoretical stars governed by these processes with what we observe in the solar neighbourhood and then make further predictions. There have been several studies that have used hydrodynamical simulations to explore r-process enrichment; however they exclusively focus on NS mergers as the primary r-process site (\citealt{Shen15}, \citealt{Voort15}). NS mergers have long been a theoretical candidate, and their existence was strongly supported by the recent gravitational wave detection GW170817 \citep{Abbott17a}, which is associated with an astronomical transient AT2017gfo \citep{Valenti17} and a short $\gamma$-ray burst GRB170817A \citep{Abbott17b}. Analysis of the kilonova suggests that it is powered by the decay of a range of r-process elements including lanthanides \citep{TanakaEtAl17} lending credit to the notion that NS mergers are a primary site for r-process production. However, it has been suggested that the timescale of NS mergers may not be short enough to explain the observations of r-process elements in extremely metal-poor stars, and thus nucleosynthesis at other sites (such as MRSNe) has also been studied \citep[e.g.,][]{Argast04}. 

In this paper, we incorporate the most recent theoretical yields for various r-process sites into our chemodynamical simulations in an attempt to constrain the dominant site. 
In Section 2, we briefly summarize the code and yield tables we use. Section 3 shows our results for isolated dwarf disc galaxies which we use to help determine parameters.
In Section 4, we show our results for a Milky Way-type galaxy, compare with observational data from the literature and include relevant discussion. Finally, in Section 5 we present our conclusions. 

\section{Code and Yields} 

\subsection{Hydrodynamical code} 

Our code is based on the smooth particle hydrodynamics (SPH) code GADGET-3 (see \citealt{Springel05} for the previous version GADGET-2).
The gravity for all particles is computed with tree method gravitational N-body dynamics and the gas elements are modelled using the entropy-conserving SPH formulation from \citet{Springel03}. A grid method is employed to govern the outermost dark matter particles that occur as a result of the initial conditions we use.
We also include the relevant physical baryonic processes from \citet{Kobayashi07}, which can be summarized as follows: 

\textit{UVB Heating}: Heating from UV background radiation is included to reproduce the observed metallicity distribution function in the solar neighbourhood (Fig 14 of \citealt{Kobayashi11b}).

\textit{Radiative Cooling}: The radiative cooling we use includes a metallicity dependency in the cooling functions calculated using the MAPPINGS III code from \citet{SutherlandDopita93}. [Fe/H] is used as the metallicity for the calculations, and the observed [$\alpha$/Fe]-[Fe/H] relation in the solar neighbourhood is assumed. In addition, kernel weighted smoothing for cooling (\citealt{Wiersma09}) is included. The impact depends on the resolution of the simulations: Kobayashi (private. comm.) found that it gives a better SFR for Milky Way simulations as in this paper, but not for cosmological simulations as in \citet{Taylor14}.  

\textit{Star Formation}: We use the star formation conditions from \citet{Katz92}, namely that there be convergent gas flow, rapid cooling and the gas is Jeans unstable. Star formation rate is related to the dynamical timescale by $\tau_{\rm sf} = \frac{1}{c}\tau_{\rm dyn}$ where we use a value of c = 0.1. The value is initially taken from \citet{Kobayashi05} and shown in \citet{Kobayashi11b} to better match abundance ratios in the Milky Way. 

Provided that a gas particle meets the above conditions, it will spawn a star particle of roughly half the mass of the initial mass of the gas particle. Star particles are a collection of stars rather than a single stellar object; we model this as a simple stellar population with a mass distribution in the newly spawned particle following the Kroupa initial mass function \citep{Kroupa08} between 0.07 and 120 $\mathrm{M_{\odot}}$. 

\textit{Feedback}: Supernovae (SNe) and stellar winds eject energy, mass and elements from the star particles back into the surrounding gas particles. These quantities are distributed to the 64 $\pm$ 2 nearest neighbour gas particles within a dynamic feedback radius and are weighted by the smoothing kernel used for the SPH. In this paper we model the energy feedback as entirely thermal, i.e., all energy from supernovae and stellar winds is distributed to the surrounding gas particles as thermal energy weighted by the kernel. The modelling of feedback is debated and we have tested kinetic feedback as in \citet{Vecchia08} and a stochastic thermal model as in \citet{Vecchia12} with our chemical enrichment method. With our code and resolution, we find that thermal feedback gives a better match with the elemental abundance distribution of the Milky Way than the other feedback methods (Haynes \& Kobayashi, in prep.) and thus we elected to use our default feedback model in this paper. 

It should be noted that no explicit sub-grid diffusion scheme for metals between gas particles (\citealt{Shen10}) is included in our model. This could change [Fe/H] but the effect on [X/Fe] for elements produced from the same sources should be less pronounced. Kobayashi (private. comm.) found that the simple diffusion equation gives too low scatter in [Fe/H] compared to observations, although it depends on the resolution and the details of the feedback modelling. The effect of additional mixing can be tested with kernel weighting \citep{Crain13} and we find that this effect should be small (see Appendix C). 

\subsection{Chemical Enrichment} 

The chemical enrichment included in our simulations is based on the model from \citet{Kobayashi04}. We detail this below with the modifications made to accommodate the additional processes from r-process nucleosynthesis. 

\textit{Core-Collapse (CC) SNe}: CCSNe includes all SNe events driven by gravitational core collapse (Type II and Ibc) SNe (SNII). We also include hypernovae (HNe), which correspond to broad-line SNe Ibc. Both yields are taken from \citep{Kobayashi11a} and govern stars between masses of 10 - 50 $M_{\odot}$; stars with $M > 50 M_{\odot}$ are assumed to collapse directly to a black hole and not to return newly synthesized metals. 
The HNe occurrence fraction is metallicity dependent: $\mathrm{f_{HNe}}$ = 0.01, 0.01, 0.23, 0.4, 0.5, and 0.5 for $Z$ = 0.05, 0.02, 0.008, 0.004, 0.001, and 0.0, respectively, and linearly interpolated for values in between \citep{Kobayashi11b}. 

\textit{Asymptotic Giant Branch (AGB) Stars}:  Stars with $\sim$ 1-8 $\mathrm{M_\odot}$ are also able to produce some metals during the AGB phase, including the heavy s-process elements. This occurs as the He shell surrounding the core pulses and allows a downflow of protons to form a $^{13}$C-rich ``pocket'' which can drive neutron production via $\alpha$ absorption (and hence s-processing). We adopt the latest nucleosynthesis yields of the AGB stars from \citet{Karakas16}.

\textit{Stellar Winds (SW)}: All stars return their envelopes containing unprocessed metals to their surroundings at the end of their lives. These metals are those that existed in the gas from which the star was formed and have their own chemical composition. 

\textit{Type Ia SNe (SNIa)}: We adopt the SNIa yields of the standard W7 model from \citet{Nomoto97}. Our progenitor model is based on single-degenerate scenario with white dwarf winds \citep{Kobayashi09}, where the rate depends both on the lifetime of the secondary stars and the metallicity of the progenitor systems, and gives good agreement with observed [$\alpha$/Fe] and [Mn/Fe] ratios. This model also gives a power-law lifetime distribution (Fig 2 of \citealt{Kobayashi09}) at $\mathrm{Z} \ge \mathrm{Z_\odot}$, very similar to the observed delay-time distribution \citep{Maoz14}. 

\textit{NS Mergers: } NS mergers provide a site for the r-process in the neutron rich dynamic ejecta created as the binary system merges. We include yield tables for NS-NS mergers (both 1.3 $\mathrm{M_{\odot}}$) from \citet{Wanajo14} and for NS-BH mergers. The NS-NS merger and NS-BH merger delay time distributions (DTD) of simple stellar populations are taken from the binary population synthesis calculations in \citet{Mennekens14} (model 2 for $Z=0.02$ and $0.002$).  The resulting yields are added to the metals released from each star particle at every time step. We also introduce a free parameter, ${f_{\rm NSM}}$, representing the fraction of stars in binary systems in our simulations. We choose an initial value of $f_{\rm NSM}=0.5$, independent of metallicity. 

\textit{MRSNe: } Rapidly rotating massive stars with strong magnetic fields may provide a potential site for r-process nucleosynthesis at the inner boundary of the accretion disc formed around the central collapsed object \citep{Cameron03}. We use the yield tables presented in \citet{Nishimura15} for a 25 $\mathrm{M_\odot}$ star (B11$\beta$1.00 model). This event could be related to HNe events, which also require rotation and magnetic fields; recent simulations of supernova explosions have not succeeded in exploding very massive stars (M $\ga$ 25 $\mathrm{M_\odot}$) \citep{Janka12}. Therefore, we replace a fraction of HNe events with MRSNe.
The number of stars with suitable conditions for MRSNe is poorly constrained so we introduce a free parameter, $f_{\rm MRSN}$, representing the fraction of stars with the correct conditions. Our initial value is $f_{\rm MRSN}=0.01$, independent of mass and metallicity, as it gives solar values at [Fe/H]$=0$ and reasonable agreement with observations. 

\textit{ECSNe: } We adopt yields from \citet{Wanajo13a} for a 8.8 $\mathrm{M_\odot}$ star and metallicity dependent ($\log Z = -4, -3, -2.4, -2.1, -1.7$) limits of the progenitor mass with the upper (8.4, 8.4, 9.0, 9.65, 9.9 $\mathrm{M_{\odot}}$) and lower (8.2, 8.25, 8.8, 9.5, 9.75 $\mathrm{M_{\odot}}$) bounds \citep{Doherty14}. For metallicities above or below the limits, we assume the rates derived from the upper and lower limits, respectively. These yields are added directly onto the ejected material from each star particle with the corresponding age. 

\textit{NUW: } Ejecta heated by neutrinos from proto-neutron stars can provide a site from r-process enrichment. The yields are taken from \citet{Wanajo13b} for stars between 13 and 40 $\mathrm{M_{\odot}}$ with a conversion between the stellar masses (13, 15, 20, 40) $\mathrm{M_{\odot}}$ and the respective NS masses (1.4, 1.6, 1.8, 2.0) $\mathrm{M_{\odot}}$ based on Kobayashi, Karakas, Lugaro (in prep). These yields are added to the SNII yields. 

Figure \ref{NucYields} shows a comparison of the yields we use. From this we show that SNII  and AGB only contribute significantly to elements up to the iron peak elements, whilst heavier elements are dominated by NS mergers and MRSNe. ECSNe produce a relatively large amount of low mass heavy elements but drop off entirely after cadmium. NUW produce elements up to and including uranium but in very low amounts compared to NS mergers and MRSNe. 

\section{Isolated Dwarf Disc Galaxies} 

\begin{figure} 
\centering 
\includegraphics[scale=0.30]{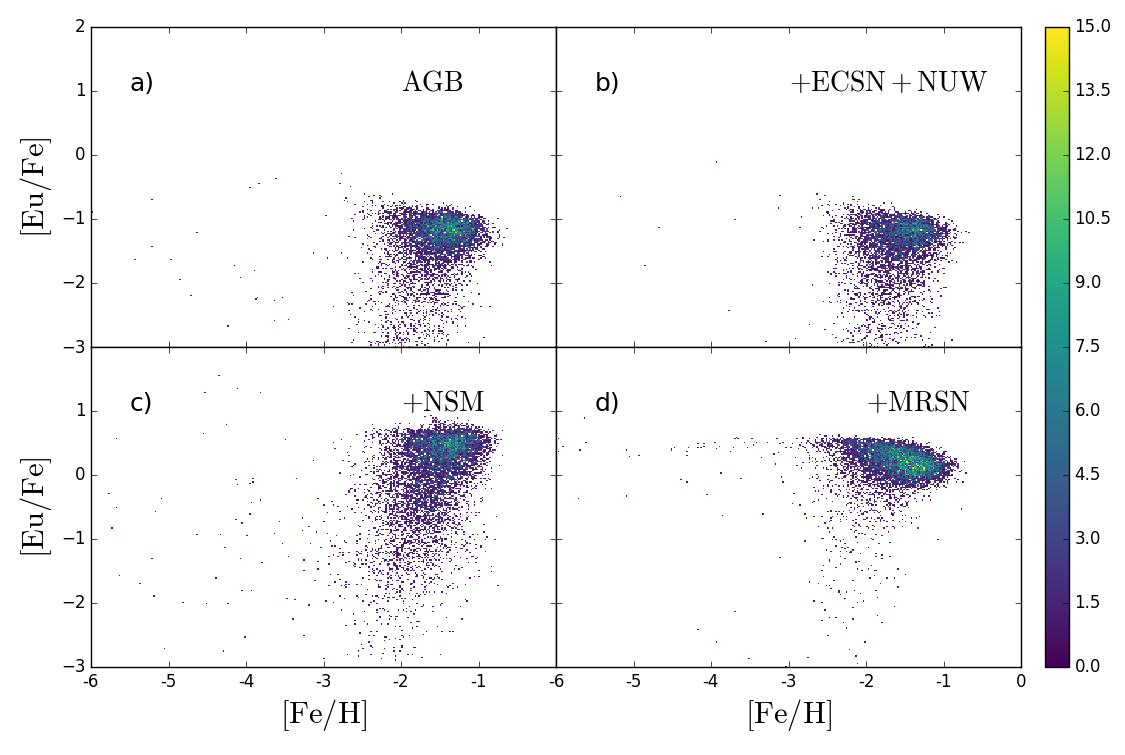} 
\caption{[Eu/Fe] plotted against [Fe/H] for the star particles in four simulations of an isolated dwarf disc galaxy: SNII + SNIa + AGB only control, NUW, NS mergers and MRSNe. The colour gradient shows the linear number of star particles per bin.} 
\label{EUFE1} 
\end{figure} 

\begin{figure} 
\centering 
\includegraphics[scale=0.30]{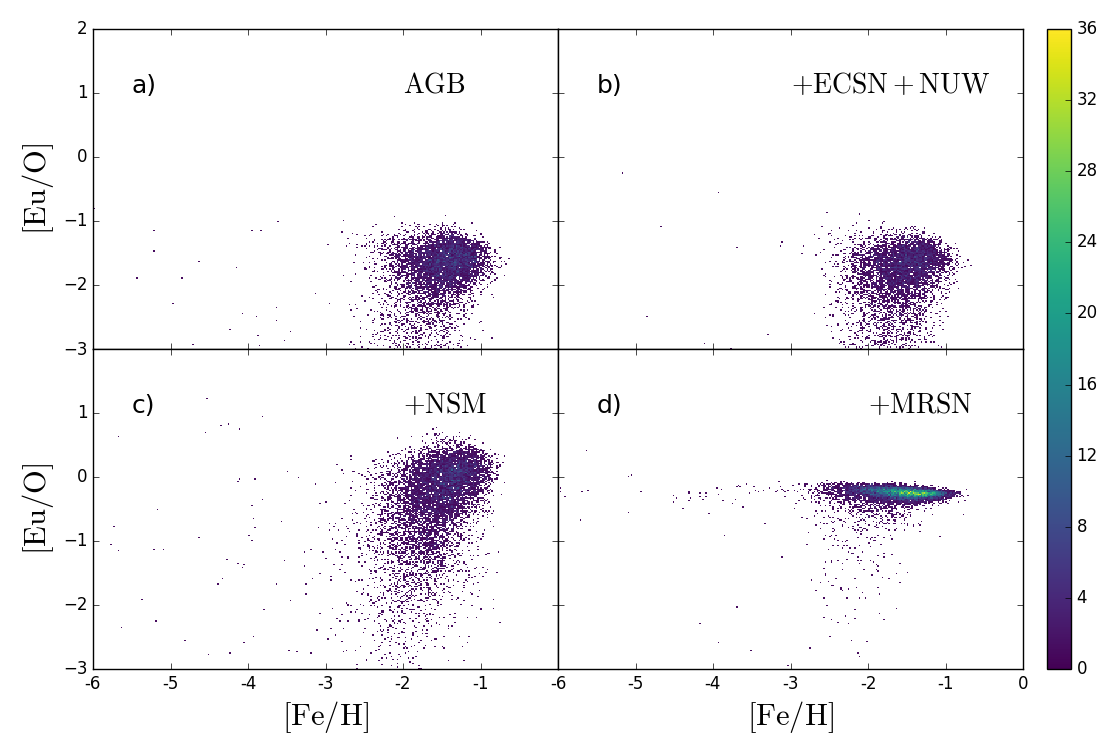} 
\caption{The same as Figure \ref{EUFE1} but showing [Eu/O] plotted against [Fe/H]. } 
\label{EUO1} 
\end{figure} 

\begin{figure*} 
\centering 
\includegraphics[scale=0.40]{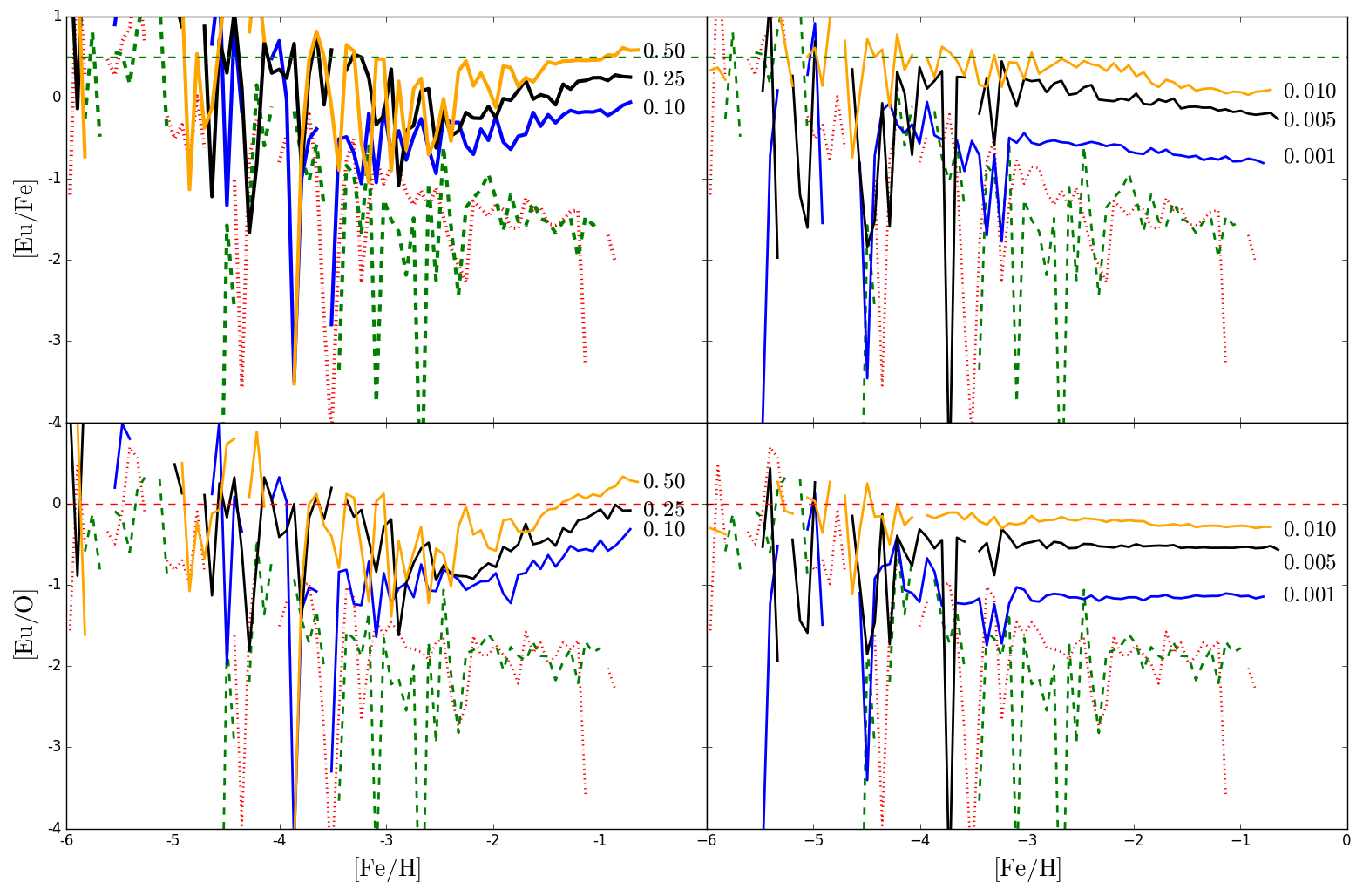} 
\caption{Median [Eu/Fe] and [Eu/O] ratios as a function of [Fe/H] for star particles in our low resolution isolated dwarf disc galaxy simulations with NS mergers (left panels) and MRSNe (right panels). Red dotted and green dashed lines show the control and ECSNe + NUW respectively. The blue, black and orange labelled lines show differing values of $f_{\rm NSM}$ (0.1, 0.25 and 0.5) and $f_{\rm MRSN}$ (0.001. 0.005, 0.01) respectively. Horizontal dashed lines show [Eu/Fe] $= 0.5$ and [Eu/O] $= 0$.} 
\label{Disc_Para} 
\end{figure*} 

In this section we present the results of our first simulations using isolated rotating gas clouds in static dark matter potentials as a test for our code and to choose parameters. The combined mass of the dark matter halo and gas is $10^{10}$ $\mathrm{h^{-1}}$ $\mathrm{M_{\odot}}$ with a 0.1 baryonic fraction and spin parameter, $\mathrm{\lambda}$ = 0.1, divided into 160 000 gas particles (see \citealt{KobayashiEtAl07}). As the simulation begins and star formation and SNe begin to take place, the cloud radiatively cools and collapses into a disc with a stellar mass of $\sim$ $\mathrm{10^{7}}$ $\mathrm{M_{\odot}}$.  

We show four different simulations. The first is effectively a control simulation with chemical enrichment just from SNII and SNIa plus the AGB yields. The remaining three simulations include the same chemical enrichment as the control but also include one of the additional r-process sites: ECSNe + NUW, MRSNe and NS mergers, respectively. We group the NUW and ECSNe into one simulation because of their relatively low yields (see Figure \ref{NucYields}). We ran many low resolution (10000 particles) simulations, changing parameters, and found that $f_{\rm MRSN}=0.01$ for MRSNe and $f_{\rm NSM}=0.5$ for NS mergers gave a good match to the solar abundance ratios (see Figure \ref{Disc_Para} below). 

Figure \ref{EUFE1} shows the [Eu/Fe] abundance ratios for the star particles in the isolated dwarf disc galaxies resulting from these simulations. Panel (a) shows the SNII + SNIa + AGB simulation (hereafter referred to as the control simulation), which depicts the base level of europium provided by the contribution from AGB stars, with no additional r-process included. As expected, given the low [Eu/Fe] in from Figure \ref{NucYields}, the NUW and ECSNe model shown in the panel (b) provides only a small boost to [Eu/Fe] at [Fe/H] $\sim -2$. Panels (c) and (d) show the NS merger and MRSNe models, respectively. Both increase the level of [Eu/Fe] substantially and to roughly the same level, with NS mergers having a much larger range in scatter between [Fe/H] $\sim -2.5$ and [Fe/H] $\sim -1$. For both models, we obtain [Eu/Fe] $\sim 0.5$ at [Fe/H] $\sim -2$, which is consistent with observations in nearby dwarf spheroidal galaxies \citep{Tolstoy09}. We note that across all models we are able to reproduce the scatter in [Eu/Fe] at low [Fe/H] that has been observed. This is due to the sporadic production of iron at low [Fe/H] prior to SNIa occurring so even relatively weak europium production can result in high [Eu/Fe]. 

In Figure \ref{EUO1} we plot [Eu/O] against [Fe/H]. We choose this particular combination as oxygen is one of the $\mathrm{\alpha}$ elements (O, Mg, Si, S, and Ca) and primarily produced in CCSNe. MRSNe are a subset of CCSNe, so oxygen should be produced at the same time, while NS mergers are independent of the occurrence of CCSNe. Additionally, by removing the Fe contribution from SNIa, we can see the relative contribution to Eu from CCSNe and neutron capture processes more directly. The panels have the same layout as Figure \ref{EUFE1}, and the panels (a) and (b) again show only a very minor increase in [Eu/O] with the addition of NUW and ECSNe. Panels (c) and (d) show a significant difference in [Eu/O] between the NS merger and MRSNe models. NS mergers keep the large scatter at [Fe/H] $\ga -3$ seen in Figure \ref{EUFE1}. However, with MRSNe the scatter is greatly reduced to form a narrow elongated peak for [Fe/H] $\ga -4$. This is due to the fact that under the MRSNe model the primary source of europium and the primary source of $\mathrm{\alpha}$ elements are both CCSNe and will therefore have similar timescales. In both models [Eu/O] approaches $\sim$ 0 at [Fe/H] $\sim$ -1. 

As mentioned earlier, in Figure \ref{Disc_Para} we summarize our parameter study result, showing the median [Eu/Fe] and [Eu/O] ratios of star particles in low resolution (10000 particles) isolated dwarf disc galaxies with different parameter choices. We note here that although these low resolution simulations display slightly decreased [X/Fe] at high [Fe/H] due to lowered SFR (see Appendix A),  this is not of concern as our primary diagnostic is the [Eu/Fe] value at [Fe/H] $\sim$ 2. Our choices of $f_{\rm MRSN}=0.01$ for MRSNe and $f_{\rm NSM}=0.5$ roughly give [Eu/Fe] $\sim$ 0.5 and [Eu/O] $\sim$ 0, therefore we use these values in the following section. With [Eu/O] ratios, we can remove the contribution from SNIa, which may be different in low-metallciity galaxies \citep{KobayashiIa15}.

\section{Milky Way Galaxies} 

In this section, we utilise cosmological zoom-in initial conditions for a Milky Way-type galaxy from the Aquila comparison project \citep{Scannapieco12} with cosmological parameters as follows: $\mathrm{H_{0}}$ = 100h = 73 km $\mathrm{s^{-1}}$ M$\mathrm{pc^{-1}}$, $\Omega_{0}$ = 0.25, $\Omega_{\Lambda}$ = 0.75, $\Omega_{B}$ = 0.04. The initial mass of each gas particle is $\mathrm{3.5 \times 10^{6} M_{\odot}}$ and the gravitational softening length is 1 $\mathrm{kpc}$ $\mathrm{h^{-1}}$. We choose these conditions because they give a galaxy with morphology, size and merger history reasonably similar to the Milky Way (see the G3-CK model in Scannapieco et al. 2012). The elemental abundance pattern, such as [$\mathrm{\alpha}$/Fe]-[Fe/H] and [Eu/Fe]-[Fe/H] relations, depends on the accretion history of galaxies \citep{Mackereth18} which could potentially be a source of systematic uncertainty. We should note, however, that our simulated galaxy displays the observed [O/Fe] bi-modality (Haynes \& Kobayashi, in prep, see Appendix B). With the same initial conditions, we run the simulation set shown in the previous section: a control galaxy (SNII + SNIa + AGB) and adding ECSNe + NUW, NSM or MRSNe. We again assume parameters of $f_{\rm MRSN}$ = 0.01 and $f_{\rm NSM}$ = 0.5 for MRSNe and NS mergers, respectively. 

\begin{figure*} 
\centering 
\includegraphics[scale=0.45]{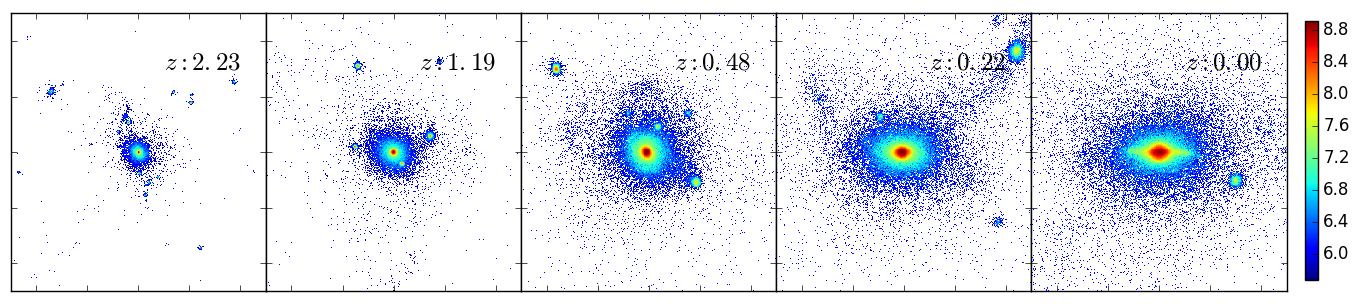} 
\caption{Projected maps of stellar mass from a comoving 100kpc box around the central galactic disc (first 5 panels) using the control simulation at a variety of redshifts (listed on each individual panel). The colour bar shows the logarithmic projected mass in $\mathrm{M_{\odot}(\frac{1}{3}kpc)^{-2}}$.} 
\label{DiscEv} 
\end{figure*} 

\begin{figure} 
\centering 
\includegraphics[scale=0.44]{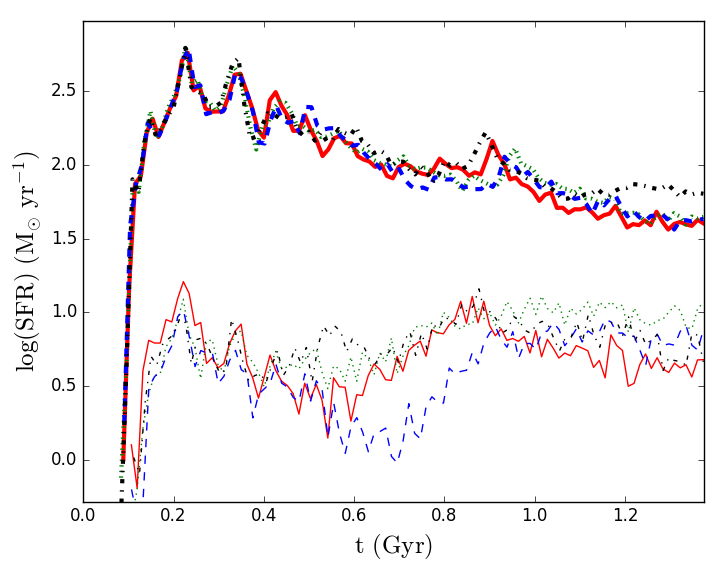} 
\caption{Star formation histories for the central 10 kpc (with height $\pm$ 2) of the galaxy (bold lines) and solar neighbourhood (thin lines). We show the star formation history for the control simulation, NUW + ECSNe, NSM and MRSNe with the red solid, green dotted, blue dashed and black dashed-dotted lines respectively.} 
\label{SFH} 
\end{figure} 

\begin{figure*} 
\centering 
\includegraphics[scale=0.38]{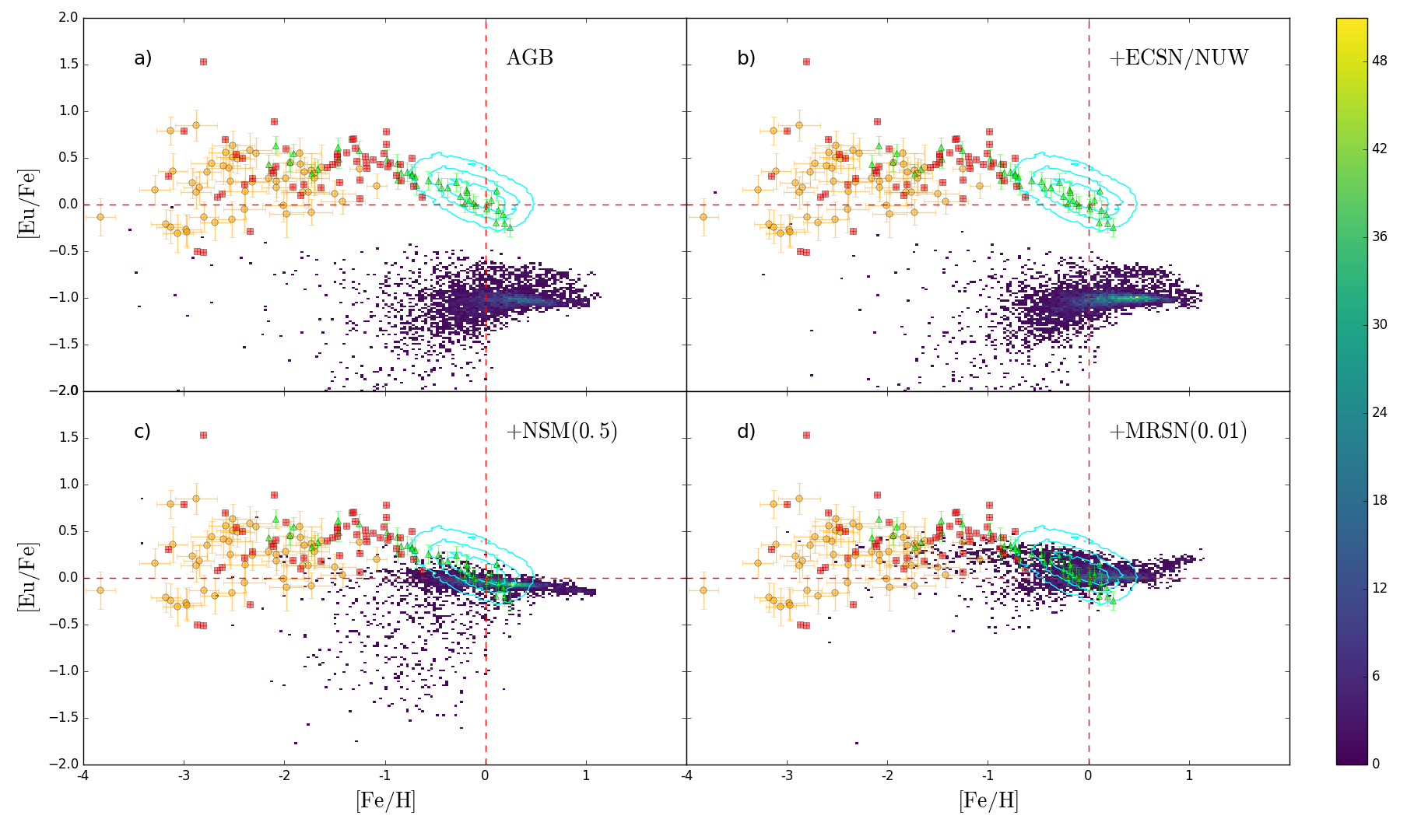} 
\caption{[Eu/Fe] plotted against [Fe/H] for the star particles in the solar neighbourhood in our Milky Way simulations at $z=0$. The panels in order show: control, ECSNe + NUW, NS mergers and MRSNe. We compare against four data sets: \citet[][red squares]{HansenEtAl16}, \citet[][orange circles]{Roederer14}, \citet[][green triangles]{ZhaoEtAl16} and \citet[][HERMES-GALAH, cyan contours]{Buder18}. The contours show 10, 50 and 100 stars per bin. The red dashed lines denote 0 for both [Eu/Fe] and [Fe/H] which we expect to lie within our data. The colour bar shows the linear number of points per bin.} 
\label{EUFEMW1} 
\end{figure*} 

\begin{figure*} 
\centering 
\includegraphics[scale=0.38]{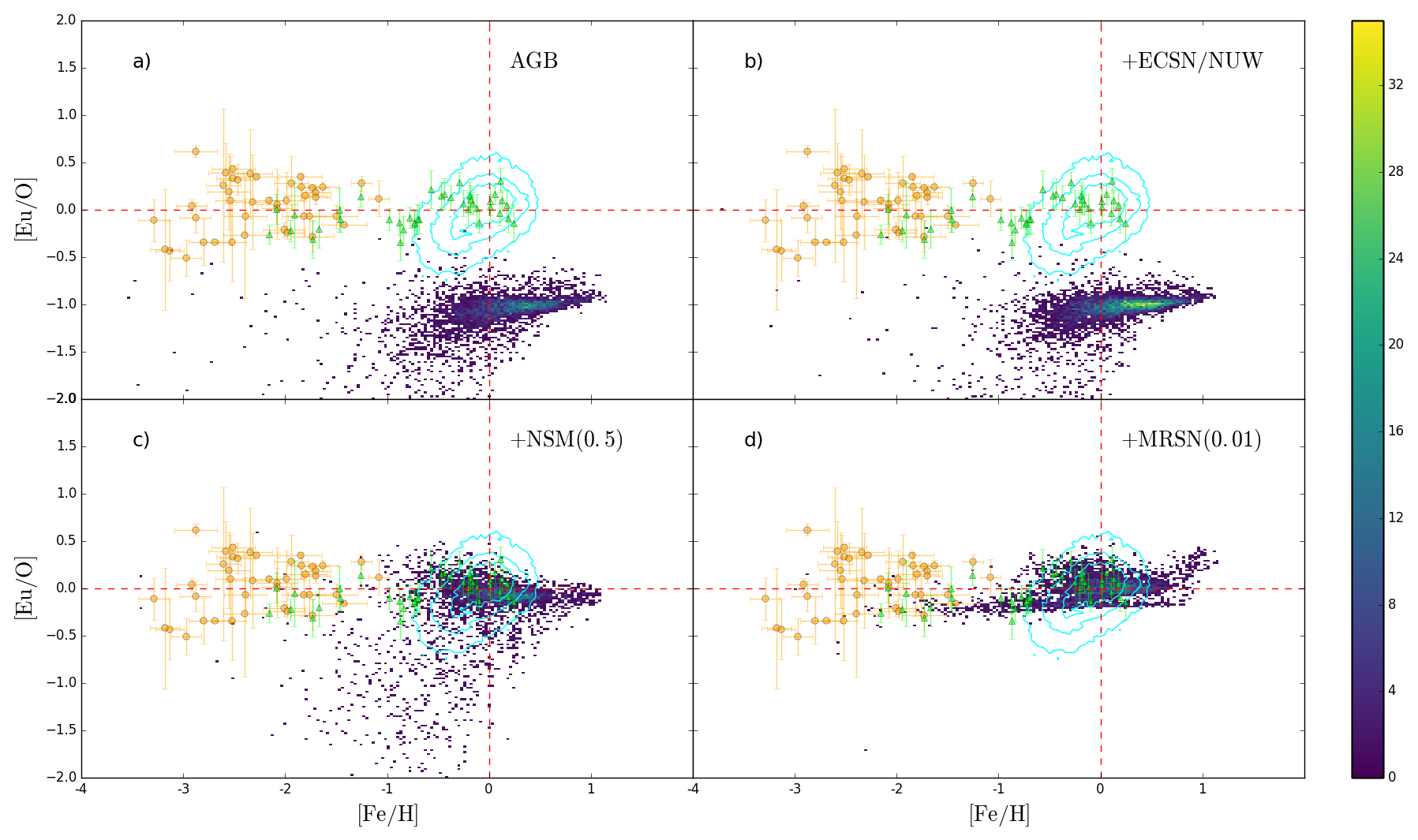} 
\caption{The same as Figure \ref{EUFEMW1} but with [Eu/O] plotted against [Fe/H]. } 
\label{EUOMW1} 
\end{figure*} 

\begin{figure} 
\centering 
\includegraphics[scale=0.35]{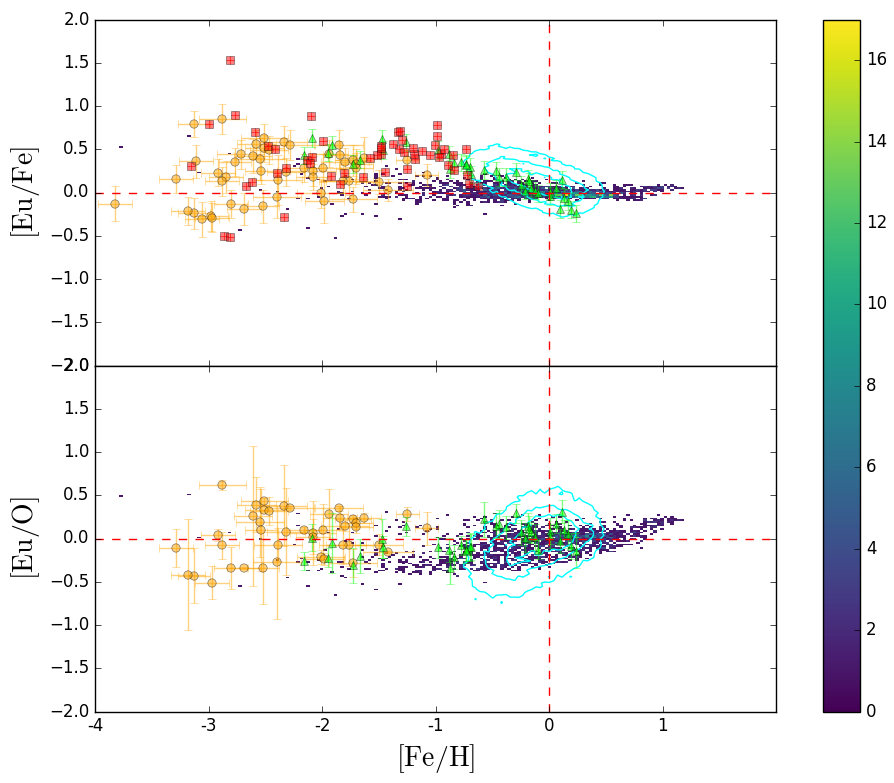} 
\caption{As Figure \ref{EUFEMW1} but showing [Eu/Fe] and [Eu/O] plotted against [Fe/H] for the combined NSM and MRSN simulation.} 
\label{EUFEOMW1} 
\end{figure} 

\begin{figure*} 
\centering 
\includegraphics[scale=0.38]{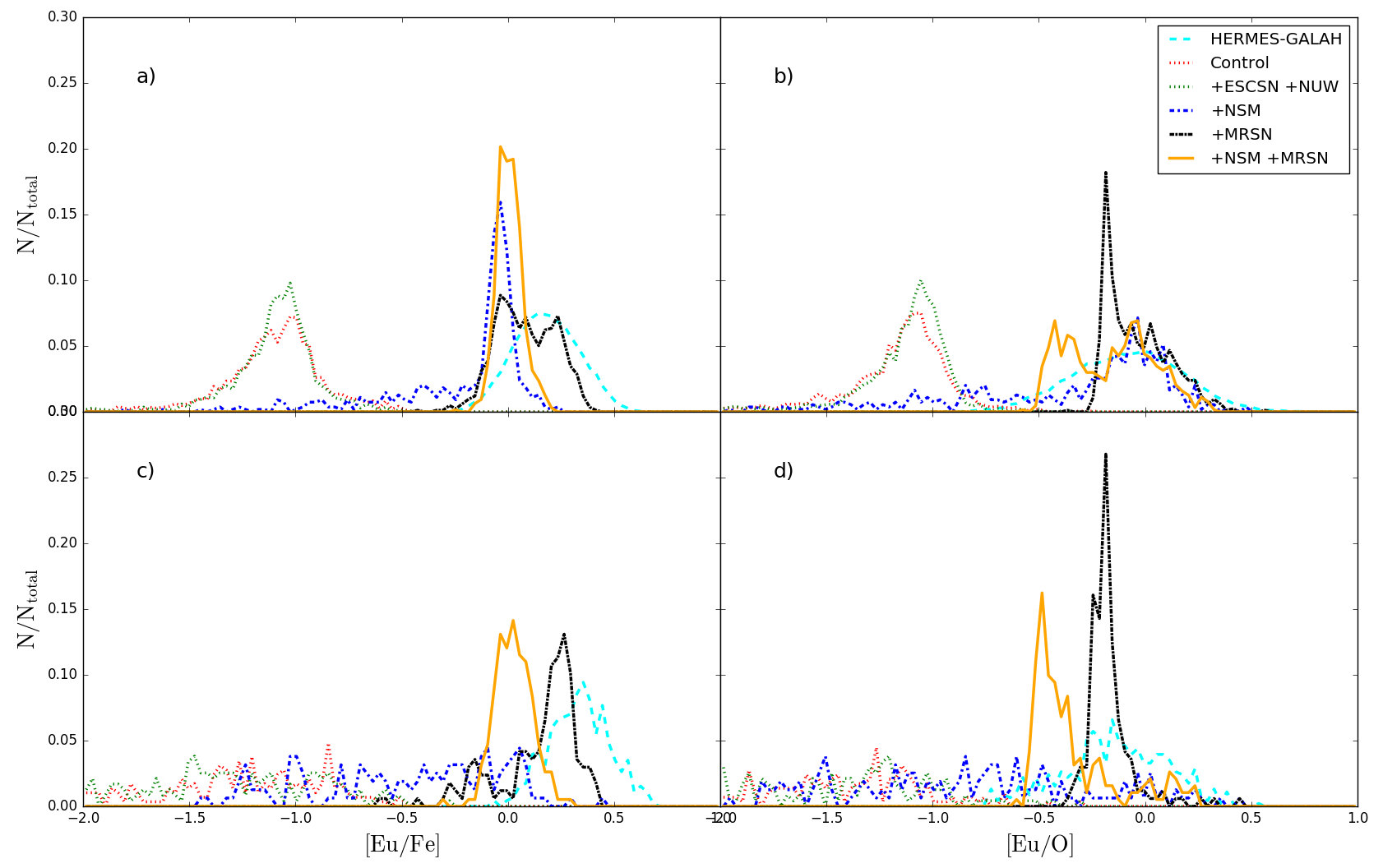} 
\caption{[Eu/Fe] distribution of star particles in the solar neighbourhood of our Milky Way simulations at $z=0$, for selected [Fe/H] ranges and normalised by the total number of points. Panels (a) and (b) show the distribution of points for $-0.75$ < [Fe/H] < $0$ for [Eu/Fe] and [Eu/O] respectively. Panels (c) and (d) show the same distributions for the range $-1.5$ < [Fe/H] < $-0.75$. The red, green, blue, and black lines show the control, ECSNe + NUW, NS mergers, and MRSNe, respectively. The orange lines show the combined model of NS mergers and MRSNe, and the cyan dashed line shows the HERMES-GALAH data in the same [Fe/H] ranges.} 
\label{EUFEHist1} 
\end{figure*} 

\begin{table*} 
\begin{tabular}{| c | c | c | c | c | c | c | c | c | c |} 
\hline 
\multicolumn{1}{| c |}{ } & \multicolumn{2}{| c |}{10 kpc disc} & \multicolumn{3}{| c |}{ } & \multicolumn{3}{| c |}{Solar Neighbourhood} & \multicolumn{1}{| c |}{ }\\
\hline 
Simulation & $\mathrm{M_{gas}}$ & $\mathrm{M_{*}}$ & $\mathrm{f_{*}}$ & $\mathrm{SFR}$ & Effective SN & $\mathrm{M_{gas}}$ & $\mathrm{M_{*}}$ & $\mathrm{f_{*}}$ & $\mathrm{SFR}$\\ 
\hline 
\multicolumn{1}{| c |}{ } & \multicolumn{2}{| c |}{$\mathrm{10^{9}}$ $\mathrm{M_{\odot}}$} & \multicolumn{1}{| c |}{ } & \multicolumn{1}{| c |}{$\mathrm{M_{\odot}}$ $\mathrm{y^{-1}}$} & \multicolumn{1}{| c |}{$\mathrm{kpc}$} & \multicolumn{2}{| c |}{$\mathrm{10^{9}}$ $\mathrm{M_{\odot}}$} & \multicolumn{1}{| c |}{ } & \multicolumn{1}{| c |}{$\mathrm{M_{\odot}}$ $\mathrm{y^{-1}}$}\\
\hline 
\hline 
Control & 10.57 & 97.19 & 0.90 & 40.12 & 5.688 - 7.438 & 1.54 & 4.13 & 0.73 & 4.7808 \\ 
ECSN + NUW & 11.64 & 100.32 & 0.90 & 44.35 & 5.883 - 7.693 & 2.91 & 5.64 & 0.66 & 11.64 \\ 
NSM & 10.60 & 96.02 & 0.90 & 43.03 & 6.500 - 8.500 & 2.23 & 3.55 & 0.61 & 8.44 \\ 
MRSN & 11.31 & 100.84 & 0.90 & 63.63 & 6.812 - 8.908 & 1.75 & 4.91 & 0.74 & 7.72 \\ 
\hline 
\label{SolarNeighbourhoodValues} 
\end{tabular} 
\caption{The properties for the simulated galaxies at $z = 0$. All masses are given in terms of $10^{9}$ $\mathrm{M_{\odot}}$ and SFR is given in $\mathrm{M_{\odot}}$ $\mathrm{y^{-1}}$. Values in the first half of the table are given for a central 10 kpc disc with height $\pm$ 2 kpc. Values in the second half correspond to the particles within the effective solar neighbourhood as given.}
\end{table*} 

Unlike the isolated dwarf disc galaxies from Section 3, N-body dynamics is used to govern dark matter behaviour to allow a central galaxy to form hierarchally in a series of mergers from cosmological initial conditions. This is easily seen in Figure \ref{DiscEv} which shows the edge-on views of the evolution of the stellar mass in 100 kpc boxes (first five panels) and a 30 kpc box (final panel) for the control simulation. The disc inclination is corrected to horizontal using the cross product vectors of the angular momentum and position vectors for each star particle. Table \ref{SolarNeighbourhoodValues} shows galactic properties in our simulations for the two regions: the central disc, defined as a 10 kpc radius disc with a height of $\pm 2$ kpc, and the solar neighbourhood, defined as the ``hollow cylinder'' with inner boundary at 6.5 kpc, outer boundary at 8.5 kpc and extending $\pm$ 1 kpc above and below the plane of the disc. Due to the nature of hydrodynamical simulations, the total stellar mass and morphology of the galaxy differs slightly between each simulation. To account for this we instead define an effective solar neighbourhood by comparing the scale widths of the simulated galaxies with the Milky Way scale width (using a figure of 2.1 $\pm$ 0.3), and adjusting the inner and outer boundaries by using the ratio between values as a simple multiplier. Finally, Figure \ref{SFH} shows the star formation rates (SFRs) for the previously defined central disc (dashed lines) and the solar neighbourhood (dotted lines) for each model. The star formation rate for the galaxy is in reasonable agreement with that from the G3-CK model in \citet{Scannapieco12}.  

Figure \ref{EUFEMW1} plots [Eu/Fe] abundances against [Fe/H] for each simulations effective solar neighbourhood. In addition we plot four observational data sets comprising derived abundances from \citet{HansenEtAl16}, \citet{Roederer14}, \citet{ZhaoEtAl16} and the HERMES-GALAH data set from \citet{Buder18}. Given the large number of observations in the HERMES-GALAH data, we condense the data into contours showing 10, 50 and 100 data points per bin (0.038 width by 0.02 height) for legibility. This results in some outlying data points not being shown towards the left hand side [Fe/H] $\sim -1$ though has little effect on the high [Fe/H] edge of the data set. We also note here that the density of observed points is unrelated to the density of simulated points outside of the HERMES-GALAH data set, as the other sets are exclusively surveying low metallicity stars. 

Panel (a) shows the control simulation with the AGB contributions but no additional r-process sites. The spread of [Eu/Fe] sits well below observational values; [Eu/Fe] is at $\sim -1$ for [Fe/H] $\ga -1.5$. Panel (b) shows the addition of both NUW and ECSNe to the AGB model. As with the isolated clouds (Section 3), the addition is insufficient to boost [Eu/Fe] to the observed levels and the overall trend remains at [Eu/Fe] $\sim -1$. This is in line with what we would expect to see from Figure \ref{NucYields}: no Eu enrichment from ECSNe and only a small contribution from NUW. 

Panels (c) and (d) show simulations using NS mergers and MRSNe, respectively. As with the addition of the NUW and ECSNe models, we have included these yields in tandem with the AGB model. Both NS mergers and MRSNe increase [Eu/Fe] to $\sim 0$ at [Fe/H] = 0, though with the caveat that both make use of a free parameter that allows the level to be adjusted reasonably freely in our simulations. 

At [Fe/H] $\la -1.5$ the simulated level and scatter of [Eu/Fe] in the MRSNe model matches the observational data much better than the NS mergers model. This is a result of delay times before NS merger events can occur, as NS binaries need to both form and coalesce. MRSNe events, on the other hand, are assumed to be a subset of 25 - 40 $\mathrm{M_{\odot}}$ HNe in our simulations and able to start producing r-process elements almost immediately after the formation of stars. Additionally, the NS merger model displays a large amount of scatter below the main trend of at [Fe/H] $\la$ 0. We already see a similar scatter at the same [Fe/H] range in the control simulation; the scatter we see in NS mergers is likely produced by AGB stars. However, this same pattern of scatter is not present in the MRSNe simulation. In this case it appears that the production of Eu from NS mergers is too slow to raise [Eu/Fe] to observed levels at [Fe/H] $< -1$ and instead slowly increases [Eu/Fe] between $-2 \la$ [Fe/H] $\la$ 0. However, MRSNe events are able to enrich the ISM prior to AGB contributions so no such scatter is seen. Although MRSNe replicate the trend in this region substantially better, it presents a flat trend above [Fe/H] $\sim -1$ where the observational data suggests a downward trend. NS mergers have a slight downward trend in this region, though not as steep as the observed data. 

Both the scatter and average [Eu/Fe] are important for constraining the r-process. SNIa produce substantial amounts of Fe which contributes to the scatter in Figure \ref{EUFEMW1}, so we show the [Eu/O] comparison in Figure \ref{EUOMW1} in order to remove the SNIa contribution. As with Figure \ref{EUFEMW1} the control and the ECSNe + NUW model and sit well below observations with the latter providing only a small peak at [Eu/O] $\sim -1$. The MRSNe model shows higher [Eu/O] at [Fe/H] $\la -1$ and provides a better match to the observational data than the NS merger model in this region. The large scatter at [Fe/H] $\la$ 0 in the NS merger model is still present as in Figure \ref{EUFEMW1}, again suggesting a slow increase in [Eu/O] due to the contribution from NS mergers. Conversely, the MRSNe overpredicts the level of [Eu/O] at [Fe/H] $\sim -0.5$, where the observational data might suggest more scatter should be exist in [Eu/O]. In the region of [Fe/H] $\ga -0.5$, both models predict a reasonably flat overall trend, however neither matches the scatter particularly well; NS mergers predict a large scatter below [Eu/O] $\sim$ 0 and MRSNe predicts a trend that is too narrow. 

At [Fe/H] $\ga$ 0,  panels (d) of Figures \ref{EUFEMW1} and \ref{EUOMW1} show predicted [Eu/Fe] and [Eu/O] ratios that increase from [Fe/H] $\sim$ 0. This upturn is the result of decreased O and Fe production as a result of MRSNe totally replacing HNe contribution. As described in Section 2, we have a metallicity dependent fraction of CCSNe that are HNe, with a constant $f_{\rm MRSN}$. At $Z=Z_{\odot}$, both fractions have a value of 0.01, which gives much larger Eu relative to O and Fe (Figure \ref{NucYields}). Although there are no stars at [Fe/H] $\ga$ 0.5 in the observational data, [Eu/O] seems to increase while [Eu/Fe] seems to decrease from [Fe/H] $\sim$ 0 to higher [Fe/H]. This may suggest some metallicity dependence of $f_{\rm MRSN}$, similar to that which is applied for HNe in our model.  

The observed data points in Figure \ref{EUOMW1} give us some insight into the timescales involved in europium production in the two models. By comparing europium to an $\mathrm{\alpha}$ element (in this case oxygen, though all $\alpha$ elements provide similar results) we compare europium production with an element produced primarily in SNII/HNe. The observed data shows a largely flat [Eu/O] trend as a function of [Fe/H]. This suggests that europium is produced primarily at the same rate (normalised to solar) as oxygen is. We assume that MRSNe events are a subset of CCSNe with conditions (e.g., rotation and magnetic fields) sufficient to produce r-process elements. Given that SNII and HNe are the primary production sites of $\mathrm{\alpha}$ elements, if MRSNe are a primary site for the r-process then most europium production would also be accompanied by $\alpha$ elements, leading to the observed flat [Eu/O] over a wide range of [Fe/H]. 

We also consider the possibility that the poor match between NS mergers and observational data could be rectified by the inclusion of MRSNe. In order to examine this, we ran a simulation that incorporated both NS mergers and MRSNe. We opted to use the same parameters for NS mergers but halved the MRSNe efficiency to $f_{\rm MRSN} = 0.005$ in an effort to prevent the overproduction of Eu at high [Fe/H]. The results of these simulations are shown in Figure \ref{EUFEOMW1}. The upper panel shows [Eu/Fe] for the combined processes and the lower panel shows [Eu/O]. The [Eu/Fe] fit with the data is slightly worse than our MRSNe model at [Fe/H] $\la -1.5$ as the smaller MRSNe contribution pulls down the overall values. The scatter seen below the observed data in the NS merger models has disappeared completely in both [Eu/Fe] and [Eu/O]. The overall trend in [Eu/Fe] is very flat and below [Fe/H] $\sim$ 0 the model underpredicts [Eu/Fe] resulting in the MRSNe model providing a better fit. 

Figure \ref{EUFEHist1} shows the [Eu/Fe] (panels a and c) and [Eu/O] (panels b and d) distributions for the ranges $-0.75$ < [Fe/H] < 0 (panels a and b) and $-1.5$ < [Fe/H] < $-0.75$ (panels c and d). The dotted lines show our four initial simulations shown in Figures 6 and 7, and the orange solid lines are for the combined NS merger and MRSNe simulation in Figure 8. The cyan dashed lines show the HERMES-GALAH observational data set.
The control simulation (red) and the combined ECSNe and NUW model (green) show low peaks at [Eu/Fe] $\sim -1$ in the panels (a) and (b) and diffuse scatter for [Eu/Fe] $\la$ $-0.5$ in the panes (c) and (d). They are both poor matches for the HERMES-GALAH data. In the panel (a) NS mergers show a narrow peak with [Eu/Fe] $\sim$ 0 as opposed to the broader peak at [Eu/Fe] $\sim$ 0.2 for the HERMES-GALAH data and in the panel (b) they do not account for the observed high values of [Eu/O] $\ga$ 0.5. Although NS mergers (blue) seem to give the best match in panel (c), they show a scatter in the range $-0.5$ $\la$ [Eu/Fe] $\la$ $-0.2$ where the HERMES-GALAH data shows no stars in panels (c) and (d). The MRSNe model (black) provides a better match in terms of peak position of [Eu/Fe] (panels a and c), and shows no scatter in the $-0.5$ $\la$ [Eu/Fe] $\la$ $-0.2$ range in the panel (a). Additionally, the MRSNe model gives fairly good fit in the panels (c) and (d). Finally, the combined NS merger and MRSNe model (orange) provides a worse fit with the data than the MRSNe model in panels (a) and (c). However, it does provide a better fit than MRSNe in panel (b).  

It is of great interest whether our conclusions depend on our numerical resolution. Although we have not explored the impact of varying resolution on [Eu/Fe] and [Eu/O], we do not think that this changes our conclusions for the following reasons. \citet{Voort15} found that increasing the resolution of their simulations increased the scatter in the [r-process/Fe] ratios at $-2.5$ < [Fe/H] < $0$ and decreased the median value at [Fe/H] < $-2$. We predict that increasing our resolution would make NS mergers less consistent with observations.  

In summary, the better [Eu/Fe] agreement is seen in the MRSNe model when compared to the NS merger model at [Fe/H] $\la -1.5$ in Figure \ref{EUFEMW1}. This suggests that at such low [Fe/H], MRSNe are the dominant r-process site. Figure \ref{EUFEHist1} provides further support for this by highlighting the increased scatter in the NS merger model. This is in agreement with the galactic chemical evolution models (GCE) with binaries from \citet{Mennekens14} which found that NS mergers could contribute strongly to the galactic r-process but could not account for the first $\sim$ 100 Myr. It is also supported by the GCE models of \citet{Cote18} which suggest that an extra r-process site could provide r-process enrichment in the early universe. Similarly large scatter in NS mergers was found in \citet{Voort15} while also providing high [r-process/Fe] at low [Fe/H] which we do not find in our simulations. This difference is likely to be caused by their adoption of an empirical r-process yield and a simple power-law DTD, which gives higher rates than our chosen DTD (derived theoretically from binary population synthesis calculations), especially at low metallicity. In our simulations the MRSN model gives much better agreement with observations, but is not a perfect match. There is good agreement in both [Eu/Fe] and [Eu/O] at [Fe/H] $\la$ 0, however above this limit it appears to slightly overpredict the amount of Eu produced. The simple combination of these two channels as in Figure 8 did not improve the matching with the observational data. Some metallicity dependence on the rates and/or yields are necessary. The NS merger delay-time we use depends on the metallicity of binary systems. The metallicity dependence on the progenitors of MRSNe has not been investigated in the simulations of the explosions in MRSNe, and is found to be very important in our chemodynamical simulations.

\section{Conclusions} 

In this paper, we have presented the results of our chemodynamical simulations including the following r-process chemical enrichment processes: ECSNe, NUW, NS mergers and MRSNe. We present results from both dwarf disc galaxies formed from rotating gas clouds and more complex cosmological zoom-in simulations that form Milky Way-type galaxies. 
We found that not only [Eu/Fe] but also [Eu/O] ratios are important to constrain the r-process sites. We then predict [Eu/Fe] and [Eu/O] trends as a function of [Fe/H] and compare them to observational data. Our findings can be summarised as follows: 

\begin{itemize} 

\item Neither ECSNe or NUW are able to adequately explain the observed europium levels. This is what we expect based on the yield tables (Fig. 1) and is shown in both our dwarf disc galaxies cloud and Milky Way simulations. 

\item Both NS mergers and MRSNe are able to produce europium in sufficient quantities (i.e. [Eu/Fe] $\sim 0$ at [Fe/H] $\sim 0$) when a reasonable parameter is chosen (binary fraction or MRSNe rate, respectively). Determining which, if either, is dominant requires comparing the predicted trends with observed data. 

\item Early r-process enrichment is likely a result of MRSNe events. This is suggested by the better match with observational data at low [Fe/H] and also physically motivated: NS events have a delay as they require both the formation of a binary system and for that system to spiral in before europium can be produced in the ejecta. This delay-time depends on the binary population synthesis calculation we used, namely, gravitational kicks at NS formation potentially forcing early mergers and nucleosynthesis. The inclusion of these kicks will be studied in our future works. 

\item The [Eu/O] ratios can provide more stringent constraints on the r-process sites. As MRSNe are a subset of CCSNe, $\alpha$ elements will be produced with a similar timescale to Eu, resulting in a flatter trend of [Eu/O] ratios. This is not the case for NS mergers. The observational data for [Eu/O] further suggests that MRSNe is the dominant mechanism at [Fe/H] $\la -1.0$. 

\item We present a combined model using both NS mergers and MRSNe for chemical enrichment. Even though the contribution from MRSNe was half that used in the solely MRSNe model it sufficient to remove the low [Eu/Fe] and [Eu/O] scatter seen in the NS merger model. However, this combined model overall did not give  better fit to observational data than the MRSNe-only model. This suggests that the metallicity dependence of the rates/yields of MRSNe and NS mergers are very important. 

\end{itemize} 

We should note that, as discussed in Section 4, [X/Fe]--[Fe/H] relations depend on the accretion history of the Milky Way Galaxy. However, it would be difficult to obtain dramatically different [Eu/(O,Fe)]--[Fe/H] relations whilst keeping the [O/Fe] distributions consistent with observations; we find a similar distribution of [O/Fe] ratios in our simulated galaxy to those observed (see Appendix B). We also note that our results depend, to some extent, on the numerical resolutions (see Appendix A) and additional mixing (see Appendix C) in chemodynamical simulations. However, neither of these effects alter our conclusion that NS mergers alone are unable to reproduce the observed [Eu/(O,Fe)]--[Fe/H] relations. 

\section*{Acknowledgements} 

This work was supported by Science and Technology Facilities Council (ST/M000958/1, ST/N504105/1). This work used the DiRAC Data Centric system at Durham University, operated by the Institute for Computational Cosmology on behalf of the STFC DiRAC HPC Facility (www.dirac.ac.uk). This equipment was funded by a BIS National E-infrastructure capital grant ST/K00042X/1, STFC capital grant ST/K00087X/1, DiRAC Operations grant ST/K003267/1 and Durham University. DiRAC is part of the National E-Infrastructure. We thank University of Hertfordshire for access to their high-performance computing facilities and we thank Volker Springel for providing the GADGET-3 code. We would like to thank Sean Ryan, Gabriele Cescutti, Fiorenzo Vincenzo, Martin Krause, Illya Mandel, Elizabeth Stanway, Stephen Smartt and Kei Kotake for fruitful discussion, and Shinya Wanajo and Nobuya Nishimura for providing the nucleosynthesis yield data. We also thank Robert Crain as the referee for constructive comments and suggestions.   

\bibliographystyle{mnras}
\bibliography{Refs}

\begin{thebibliography}{}
\makeatletter
\relax
\def\mn@urlcharsother{\let\do\@makeother \do\$\do\&\do\#\do\^\do\_\do\%\do\~}
\def\mn@doi{\begingroup\mn@urlcharsother \@ifnextchar [ {\mn@doi@}
  {\mn@doi@[]}}
\def\mn@doi@[#1]#2{\def\@tempa{#1}\ifx\@tempa\@empty \href
  {http://dx.doi.org/#2} {doi:#2}\else \href {http://dx.doi.org/#2} {#1}\fi
  \endgroup}
\def\mn@eprint#1#2{\mn@eprint@#1:#2::\@nil}
\def\mn@eprint@arXiv#1{\href {http://arxiv.org/abs/#1} {{\tt arXiv:#1}}}
\def\mn@eprint@dblp#1{\href {http://dblp.uni-trier.de/rec/bibtex/#1.xml}
  {dblp:#1}}
\def\mn@eprint@#1:#2:#3:#4\@nil{\def\@tempa {#1}\def\@tempb {#2}\def\@tempc
  {#3}\ifx \@tempc \@empty \let \@tempc \@tempb \let \@tempb \@tempa \fi \ifx
  \@tempb \@empty \def\@tempb {arXiv}\fi \@ifundefined
  {mn@eprint@\@tempb}{\@tempb:\@tempc}{\expandafter \expandafter \csname
  mn@eprint@\@tempb\endcsname \expandafter{\@tempc}}}

\bibitem[\protect\citeauthoryear{{Abbott} et~al.,}{{Abbott}
  et~al.}{2017a}]{Abbott17a}
{Abbott} B.~P.,  et~al., 2017a, \mn@doi [Physical Review Letters]
  {10.1103/PhysRevLett.119.161101}, \href
  {http://adsabs.harvard.edu/abs/2017PhRvL.119p1101A} {119, 161101}

\bibitem[\protect\citeauthoryear{{Abbott} et~al.,}{{Abbott}
  et~al.}{2017b}]{Abbott17b}
{Abbott} B.~P.,  et~al., 2017b, \mn@doi [\apjl] {10.3847/2041-8213/aa920c},
  \href {http://adsabs.harvard.edu/abs/2017ApJ...848L..13A} {848, L13}

\bibitem[\protect\citeauthoryear{{Argast}, {Samland}, {Thielemann}  \&
  {Qian}}{{Argast} et~al.}{2004}]{Argast04}
{Argast} D.,  {Samland} M.,  {Thielemann} F.-K.,   {Qian} Y.-Z.,  2004, \mn@doi
  [\aap] {10.1051/0004-6361:20034265}, \href
  {http://adsabs.harvard.edu/abs/2004A%26A...416..997A} {416, 997}

\bibitem[\protect\citeauthoryear{{Bisterzo}, {Gallino}, {Straniero},
  {Cristallo}  \& {K{\"a}ppeler}}{{Bisterzo} et~al.}{2011}]{Kaeppeler11}
{Bisterzo} S.,  {Gallino} R.,  {Straniero} O.,  {Cristallo} S.,
  {K{\"a}ppeler} F.,  2011, \mn@doi [\mnras]
  {10.1111/j.1365-2966.2011.19484.x}, \href
  {http://adsabs.harvard.edu/abs/2011MNRAS.418..284B} {418, 284}

\bibitem[\protect\citeauthoryear{{Buder} et~al.,}{{Buder}
  et~al.}{2018}]{Buder18}
{Buder} S.,  et~al., 2018, \mn@doi [\mnras] {10.1093/mnras/sty1281}, \href
  {http://adsabs.harvard.edu/abs/2018MNRAS.tmp.1218B} {}

\bibitem[\protect\citeauthoryear{{Cameron}}{{Cameron}}{2003}]{Cameron03}
{Cameron} A.~G.~W.,  2003, \mn@doi [\apj] {10.1086/368110}, \href
  {http://adsabs.harvard.edu/abs/2003ApJ...587..327C} {587, 327}

\bibitem[\protect\citeauthoryear{{Cescutti}, {Romano}, {Matteucci}, {Chiappini}
   \& {Hirschi}}{{Cescutti} et~al.}{2015}]{Cescutti15}
{Cescutti} G.,  {Romano} D.,  {Matteucci} F.,  {Chiappini} C.,   {Hirschi} R.,
  2015, \mn@doi [\aap] {10.1051/0004-6361/201525698}, \href
  {http://adsabs.harvard.edu/abs/2015A%26A...577A.139C} {577, A139}

\bibitem[\protect\citeauthoryear{{C{\^o}t{\'e}} et~al.,}{{C{\^o}t{\'e}}
  et~al.}{2018}]{Cote18}
{C{\^o}t{\'e}} B.,  et~al., 2018, preprint, \href
  {http://adsabs.harvard.edu/abs/2018arXiv180903525C} {} (\mn@eprint {arXiv}
  {1809.03525})

\bibitem[\protect\citeauthoryear{{Crain}, {McCarthy}, {Schaye}, {Theuns}  \&
  {Frenk}}{{Crain} et~al.}{2013}]{Crain13}
{Crain} R.~A.,  {McCarthy} I.~G.,  {Schaye} J.,  {Theuns} T.,   {Frenk} C.~S.,
  2013, \mn@doi [\mnras] {10.1093/mnras/stt649}, \href
  {http://adsabs.harvard.edu/abs/2013MNRAS.432.3005C} {432, 3005}

\bibitem[\protect\citeauthoryear{{Dalla Vecchia} \& {Schaye}}{{Dalla Vecchia}
  \& {Schaye}}{2008}]{Vecchia08}
{Dalla Vecchia} C.,  {Schaye} J.,  2008, \mn@doi [\mnras]
  {10.1111/j.1365-2966.2008.13322.x}, \href
  {http://adsabs.harvard.edu/abs/2008MNRAS.387.1431D} {387, 1431}

\bibitem[\protect\citeauthoryear{{Dalla Vecchia} \& {Schaye}}{{Dalla Vecchia}
  \& {Schaye}}{2012}]{Vecchia12}
{Dalla Vecchia} C.,  {Schaye} J.,  2012, \mn@doi [\mnras]
  {10.1111/j.1365-2966.2012.21704.x}, \href
  {http://adsabs.harvard.edu/abs/2012MNRAS.426..140D} {426, 140}

\bibitem[\protect\citeauthoryear{{Doherty}, {Gil-Pons}, {Siess}, {Lattanzio}
  \& {Lau}}{{Doherty} et~al.}{2015}]{Doherty14}
{Doherty} C.~L.,  {Gil-Pons} P.,  {Siess} L.,  {Lattanzio} J.~C.,   {Lau}
  H.~H.~B.,  2015, \mn@doi [\mnras] {10.1093/mnras/stu2180}, \href
  {http://adsabs.harvard.edu/abs/2015MNRAS.446.2599D} {446, 2599}

\bibitem[\protect\citeauthoryear{{Freiburghaus}, {Rosswog}  \&
  {Thielemann}}{{Freiburghaus} et~al.}{1999}]{FreiburghausEtAl99}
{Freiburghaus} C.,  {Rosswog} S.,   {Thielemann} F.-K.,  1999, \mn@doi [\apjl]
  {10.1086/312343}, \href {http://adsabs.harvard.edu/abs/1999ApJ...525L.121F}
  {525, L121}

\bibitem[\protect\citeauthoryear{{Frischknecht} et~al.,}{{Frischknecht}
  et~al.}{2016}]{Frischknecht16}
{Frischknecht} U.,  et~al., 2016, \mn@doi [\mnras] {10.1093/mnras/stv2723},
  \href {http://adsabs.harvard.edu/abs/2016MNRAS.456.1803F} {456, 1803}

\bibitem[\protect\citeauthoryear{{Hansen} et~al.,}{{Hansen}
  et~al.}{2016}]{HansenEtAl16}
{Hansen} C.~J.,  et~al., 2016, \mn@doi [\aap] {10.1051/0004-6361/201526895},
  \href {http://adsabs.harvard.edu/abs/2016A%26A...588A..37H} {588, A37}

\bibitem[\protect\citeauthoryear{{Janka}}{{Janka}}{2012}]{Janka12}
{Janka} H.-T.,  2012, \mn@doi [Annual Review of Nuclear and Particle Science]
  {10.1146/annurev-nucl-102711-094901}, \href
  {http://adsabs.harvard.edu/abs/2012ARNPS..62..407J} {62, 407}

\bibitem[\protect\citeauthoryear{{Karakas} \& {Lugaro}}{{Karakas} \&
  {Lugaro}}{2016}]{Karakas16}
{Karakas} A.~I.,  {Lugaro} M.,  2016, \mn@doi [\apj]
  {10.3847/0004-637X/825/1/26}, \href
  {http://adsabs.harvard.edu/abs/2016ApJ...825...26K} {825, 26}

\bibitem[\protect\citeauthoryear{{Katz}}{{Katz}}{1992}]{Katz92}
{Katz} N.,  1992, APJ, 391, 502

\bibitem[\protect\citeauthoryear{{Kobayashi}}{{Kobayashi}}{2004}]{Kobayashi04}
{Kobayashi} C.,  2004, \mn@doi [mnras] {10.1111/j.1365-2966.2004.07258.x},
  \href {http://adsabs.harvard.edu/abs/2004MNRAS.347..740K} {347, 740}

\bibitem[\protect\citeauthoryear{{Kobayashi}}{{Kobayashi}}{2005}]{Kobayashi05}
{Kobayashi} C.,  2005, \mn@doi [\mnras] {10.1111/j.1365-2966.2005.09248.x},
  \href {http://adsabs.harvard.edu/abs/2005MNRAS.361.1216K} {361, 1216}

\bibitem[\protect\citeauthoryear{{Kobayashi}}{{Kobayashi}}{2007}]{Kobayashi07}
{Kobayashi} C.,  2007, in {Emsellem} E.,  {Wozniak} H.,  {Massacrier} G.,
  {Gonzalez} J.-F.,  {Devriendt} J.,   {Champavert} N.,  eds,  EAS Publications
  Series Vol. 24, EAS Publications Series. pp 245--250 (\mn@eprint {}
  {astro-ph/0611805}), \mn@doi{10.1051/eas:2007033}

\bibitem[\protect\citeauthoryear{{Kobayashi} \& {Nakasato}}{{Kobayashi} \&
  {Nakasato}}{2011}]{Kobayashi11b}
{Kobayashi} C.,  {Nakasato} N.,  2011, \mn@doi [\apj]
  {10.1088/0004-637X/729/1/16}, \href
  {http://adsabs.harvard.edu/abs/2011ApJ...729...16K} {729, 16}

\bibitem[\protect\citeauthoryear{{Kobayashi} \& {Nomoto}}{{Kobayashi} \&
  {Nomoto}}{2009}]{Kobayashi09}
{Kobayashi} C.,  {Nomoto} K.,  2009, \mn@doi [\apj]
  {10.1088/0004-637X/707/2/1466}, \href
  {http://adsabs.harvard.edu/abs/2009ApJ...707.1466K} {707, 1466}

\bibitem[\protect\citeauthoryear{{Kobayashi}, {Springel}  \&
  {White}}{{Kobayashi} et~al.}{2007}]{KobayashiEtAl07}
{Kobayashi} C.,  {Springel} V.,   {White} S.~D.~M.,  2007, \mn@doi [mnras]
  {10.1111/j.1365-2966.2007.11555.x}, \href
  {http://adsabs.harvard.edu/abs/2007MNRAS.376.1465K} {376, 1465}

\bibitem[\protect\citeauthoryear{{Kobayashi}, {Karakas}  \&
  {Umeda}}{{Kobayashi} et~al.}{2011}]{Kobayashi11a}
{Kobayashi} C.,  {Karakas} A.~I.,   {Umeda} H.,  2011, \mn@doi [mnras]
  {10.1111/j.1365-2966.2011.18621.x}, \href
  {http://adsabs.harvard.edu/abs/2011MNRAS.414.3231K} {414, 3231}

\bibitem[\protect\citeauthoryear{{Kobayashi}, {Nomoto}  \&
  {Hachisu}}{{Kobayashi} et~al.}{2015}]{KobayashiIa15}
{Kobayashi} C.,  {Nomoto} K.,   {Hachisu} I.,  2015, \mn@doi [\apjl]
  {10.1088/2041-8205/804/1/L24}, \href
  {http://adsabs.harvard.edu/abs/2015ApJ...804L..24K} {804, L24}

\bibitem[\protect\citeauthoryear{{Kroupa}}{{Kroupa}}{2008}]{Kroupa08}
{Kroupa} P.,  2008, in {Knapen} J.~H.,  {Mahoney} T.~J.,   {Vazdekis} A.,  eds,
   Astronomical Society of the Pacific Conference Series Vol. 390, Pathways
  Through an Eclectic Universe. p.~3 (\mn@eprint {arXiv} {0708.1164})

\bibitem[\protect\citeauthoryear{{Lattimer} \& {Schramm}}{{Lattimer} \&
  {Schramm}}{1974}]{Lattimer74}
{Lattimer} J.~M.,  {Schramm} D.~N.,  1974, \mn@doi [\apjl] {10.1086/181612},
  \href {http://adsabs.harvard.edu/abs/1974ApJ...192L.145L} {192, L145}

\bibitem[\protect\citeauthoryear{{Mackereth}, {Crain}, {Schiavon}, {Schaye},
  {Theuns}  \& {Schaller}}{{Mackereth} et~al.}{2018}]{Mackereth18}
{Mackereth} J.~T.,  {Crain} R.~A.,  {Schiavon} R.~P.,  {Schaye} J.,  {Theuns}
  T.,   {Schaller} M.,  2018, \mn@doi [\mnras] {10.1093/mnras/sty972}, \href
  {http://adsabs.harvard.edu/abs/2018MNRAS.477.5072M} {477, 5072}

\bibitem[\protect\citeauthoryear{{Maoz}, {Mannucci}  \& {Nelemans}}{{Maoz}
  et~al.}{2014}]{Maoz14}
{Maoz} D.,  {Mannucci} F.,   {Nelemans} G.,  2014, \mn@doi [\araa]
  {10.1146/annurev-astro-082812-141031}, \href
  {http://adsabs.harvard.edu/abs/2014ARA%26A..52..107M} {52, 107}

\bibitem[\protect\citeauthoryear{{Mennekens} \& {Vanbeveren}}{{Mennekens} \&
  {Vanbeveren}}{2014}]{Mennekens14}
{Mennekens} N.,  {Vanbeveren} D.,  2014, \mn@doi [\aap]
  {10.1051/0004-6361/201322198}, \href
  {http://adsabs.harvard.edu/abs/2014A%26A...564A.134M} {564, A134}

\bibitem[\protect\citeauthoryear{{Nishimura}, {Takiwaki}  \&
  {Thielemann}}{{Nishimura} et~al.}{2015}]{Nishimura15}
{Nishimura} N.,  {Takiwaki} T.,   {Thielemann} F.-K.,  2015, \mn@doi [\apj]
  {10.1088/0004-637X/810/2/109}, \href
  {http://adsabs.harvard.edu/abs/2015ApJ...810..109N} {810, 109}

\bibitem[\protect\citeauthoryear{{Nomoto}, {Iwamoto}, {Nakasato}, {Thielemann},
  {Brachwitz}, {Tsujimoto}, {Kubo}  \& {Kishimoto}}{{Nomoto}
  et~al.}{1997}]{Nomoto97}
{Nomoto} K.,  {Iwamoto} K.,  {Nakasato} N.,  {Thielemann} F.-K.,  {Brachwitz}
  F.,  {Tsujimoto} T.,  {Kubo} Y.,   {Kishimoto} N.,  1997, \mn@doi [Nuclear
  Physics A] {10.1016/S0375-9474(97)00291-1}, \href
  {http://adsabs.harvard.edu/abs/1997NuPhA.621..467N} {621, 467}

\bibitem[\protect\citeauthoryear{{Ojima}, {Ishimaru}, {Wanajo}  \&
  {Prantzos}}{{Ojima} et~al.}{2017}]{Ishimaru17}
{Ojima} T.,  {Ishimaru} Y.,  {Wanajo} S.,   {Prantzos} N.,  2017, in {Kubono}
  S.,  {Kajino} T.,  {Nishimura} S.,  {Isobe} T.,  {Nagataki} S.,  {Shima} T.,
   {Takeda} Y.,  eds, 14th International Symposium on Nuclei in the Cosmos
  (NIC2016). p. 020208, \mn@doi{10.7566/JPSCP.14.020208}

\bibitem[\protect\citeauthoryear{{Prantzos}, {Hashimoto}  \&
  {Nomoto}}{{Prantzos} et~al.}{1990}]{PrantzosEtAl90}
{Prantzos} N.,  {Hashimoto} M.,   {Nomoto} K.,  1990, \aap, \href
  {http://adsabs.harvard.edu/abs/1990A%26A...234..211P} {234, 211}

\bibitem[\protect\citeauthoryear{{Roederer}, {Preston}, {Thompson}, {Shectman},
  {Sneden}, {Burley}  \& {Kelson}}{{Roederer} et~al.}{2014}]{Roederer14}
{Roederer} I.~U.,  {Preston} G.~W.,  {Thompson} I.~B.,  {Shectman} S.~A.,
  {Sneden} C.,  {Burley} G.~S.,   {Kelson} D.~D.,  2014, \mn@doi [\aj]
  {10.1088/0004-6256/147/6/136}, \href
  {http://adsabs.harvard.edu/abs/2014AJ....147..136R} {147, 136}

\bibitem[\protect\citeauthoryear{{Scannapieco} et~al.,}{{Scannapieco}
  et~al.}{2012}]{Scannapieco12}
{Scannapieco} C.,  et~al., 2012, \mn@doi [\mnras]
  {10.1111/j.1365-2966.2012.20993.x}, \href
  {http://adsabs.harvard.edu/abs/2012MNRAS.423.1726S} {423, 1726}

\bibitem[\protect\citeauthoryear{{Shen}, {Wadsley}  \& {Stinson}}{{Shen}
  et~al.}{2010}]{Shen10}
{Shen} S.,  {Wadsley} J.,   {Stinson} G.,  2010, \mn@doi [\mnras]
  {10.1111/j.1365-2966.2010.17047.x}, \href
  {http://adsabs.harvard.edu/abs/2010MNRAS.407.1581S} {407, 1581}

\bibitem[\protect\citeauthoryear{{Shen}, {Cooke}, {Ramirez-Ruiz}, {Madau},
  {Mayer}  \& {Guedes}}{{Shen} et~al.}{2015}]{Shen15}
{Shen} S.,  {Cooke} R.~J.,  {Ramirez-Ruiz} E.,  {Madau} P.,  {Mayer} L.,
  {Guedes} J.,  2015, \mn@doi [\apj] {10.1088/0004-637X/807/2/115}, \href
  {http://adsabs.harvard.edu/abs/2015ApJ...807..115S} {807, 115}

\bibitem[\protect\citeauthoryear{{Smith} \& {Lambert}}{{Smith} \&
  {Lambert}}{1990}]{SmithLambert90}
{Smith} V.~V.,  {Lambert} D.~L.,  1990, \mn@doi [\apjs] {10.1086/191421}, \href
  {http://adsabs.harvard.edu/abs/1990ApJS...72..387S} {72, 387}

\bibitem[\protect\citeauthoryear{{Springel}}{{Springel}}{2005}]{Springel05}
{Springel} V.,  2005, \mn@doi [mnras] {10.1111/j.1365-2966.2005.09655.x}, \href
  {http://adsabs.harvard.edu/abs/2005MNRAS.364.1105S} {364, 1105}

\bibitem[\protect\citeauthoryear{{Springel} \& {Hernquist}}{{Springel} \&
  {Hernquist}}{2003}]{Springel03}
{Springel} V.,  {Hernquist} L.,  2003, \mn@doi [\mnras]
  {10.1046/j.1365-8711.2003.06206.x}, \href
  {http://adsabs.harvard.edu/abs/2003MNRAS.339..289S} {339, 289}

\bibitem[\protect\citeauthoryear{{Sutherland} \& {Dopita}}{{Sutherland} \&
  {Dopita}}{1993}]{SutherlandDopita93}
{Sutherland} R.,  {Dopita} M.,  1993, APJ, 88, 253

\bibitem[\protect\citeauthoryear{{Tanaka} et~al.,}{{Tanaka}
  et~al.}{2017}]{TanakaEtAl17}
{Tanaka} M.,  et~al., 2017, \mn@doi [\pasj] {10.1093/pasj/psx121}, \href
  {http://adsabs.harvard.edu/abs/2017PASJ...69..102T} {69, 102}

\bibitem[\protect\citeauthoryear{{Taylor} \& {Kobayashi}}{{Taylor} \&
  {Kobayashi}}{2014}]{Taylor14}
{Taylor} P.,  {Kobayashi} C.,  2014, \mn@doi [\mnras] {10.1093/mnras/stu983},
  \href {http://adsabs.harvard.edu/abs/2014MNRAS.442.2751T} {442, 2751}

\bibitem[\protect\citeauthoryear{{Tolstoy}, {Hill}  \& {Tosi}}{{Tolstoy}
  et~al.}{2009}]{Tolstoy09}
{Tolstoy} E.,  {Hill} V.,   {Tosi} M.,  2009, \mn@doi [\araa]
  {10.1146/annurev-astro-082708-101650}, \href
  {http://adsabs.harvard.edu/abs/2009ARA%26A..47..371T} {47, 371}

\bibitem[\protect\citeauthoryear{{Valenti} et~al.,}{{Valenti}
  et~al.}{2017}]{Valenti17}
{Valenti} S.,  et~al., 2017, \mn@doi [\apjl] {10.3847/2041-8213/aa8edf}, \href
  {http://adsabs.harvard.edu/abs/2017ApJ...848L..24V} {848, L24}

\bibitem[\protect\citeauthoryear{{Wanajo}}{{Wanajo}}{2013}]{Wanajo13b}
{Wanajo} S.,  2013, \mn@doi [\apjl] {10.1088/2041-8205/770/2/L22}, \href
  {http://adsabs.harvard.edu/abs/2013ApJ...770L..22W} {770, L22}

\bibitem[\protect\citeauthoryear{{Wanajo}, {Kajino}, {Mathews}  \&
  {Otsuki}}{{Wanajo} et~al.}{2001}]{Wanajo01}
{Wanajo} S.,  {Kajino} T.,  {Mathews} G.~J.,   {Otsuki} K.,  2001, \mn@doi
  [\apj] {10.1086/321339}, \href
  {http://adsabs.harvard.edu/abs/2001ApJ...554..578W} {554, 578}

\bibitem[\protect\citeauthoryear{{Wanajo}, {Janka}  \& {M{\"u}ller}}{{Wanajo}
  et~al.}{2011}]{Wanajo11}
{Wanajo} S.,  {Janka} H.-T.,   {M{\"u}ller} B.,  2011, \mn@doi [\apjl]
  {10.1088/2041-8205/726/2/L15}, \href
  {http://adsabs.harvard.edu/abs/2011ApJ...726L..15W} {726, L15}

\bibitem[\protect\citeauthoryear{{Wanajo}, {Janka}  \& {M{\"u}ller}}{{Wanajo}
  et~al.}{2013}]{Wanajo13a}
{Wanajo} S.,  {Janka} H.-T.,   {M{\"u}ller} B.,  2013, \mn@doi [\apjl]
  {10.1088/2041-8205/767/2/L26}, \href
  {http://adsabs.harvard.edu/abs/2013ApJ...767L..26W} {767, L26}

\bibitem[\protect\citeauthoryear{{Wanajo}, {Sekiguchi}, {Nishimura}, {Kiuchi},
  {Kyutoku}  \& {Shibata}}{{Wanajo} et~al.}{2014}]{Wanajo14}
{Wanajo} S.,  {Sekiguchi} Y.,  {Nishimura} N.,  {Kiuchi} K.,  {Kyutoku} K.,
  {Shibata} M.,  2014, \mn@doi [\apjl] {10.1088/2041-8205/789/2/L39}, \href
  {http://adsabs.harvard.edu/abs/2014ApJ...789L..39W} {789, L39}

\bibitem[\protect\citeauthoryear{{Wehmeyer}, {Pignatari}  \&
  {Thielemann}}{{Wehmeyer} et~al.}{2015}]{Wehmeyer15}
{Wehmeyer} B.,  {Pignatari} M.,   {Thielemann} F.-K.,  2015, \mn@doi [\mnras]
  {10.1093/mnras/stv1352}, \href
  {http://adsabs.harvard.edu/abs/2015MNRAS.452.1970W} {452, 1970}

\bibitem[\protect\citeauthoryear{{Wiersma}, {Schaye}, {Theuns}, {Dalla Vecchia}
   \& {Tornatore}}{{Wiersma} et~al.}{2009}]{Wiersma09}
{Wiersma} R.~P.~C.,  {Schaye} J.,  {Theuns} T.,  {Dalla Vecchia} C.,
  {Tornatore} L.,  2009, \mn@doi [\mnras] {10.1111/j.1365-2966.2009.15331.x},
  \href {http://adsabs.harvard.edu/abs/2009MNRAS.399..574W} {399, 574}

\bibitem[\protect\citeauthoryear{{Zhao} et~al.,}{{Zhao}
  et~al.}{2016}]{ZhaoEtAl16}
{Zhao} G.,  et~al., 2016, \mn@doi [\apj] {10.3847/1538-4357/833/2/225}, \href
  {http://adsabs.harvard.edu/abs/2016ApJ...833..225Z} {833, 225}

\bibitem[\protect\citeauthoryear{{van de Voort}, {Quataert}, {Hopkins},
  {Kere{\v s}}  \& {Faucher-Gigu{\`e}re}}{{van de Voort}
  et~al.}{2015}]{Voort15}
{van de Voort} F.,  {Quataert} E.,  {Hopkins} P.~F.,  {Kere{\v s}} D.,
  {Faucher-Gigu{\`e}re} C.-A.,  2015, \mn@doi [\mnras] {10.1093/mnras/stu2404},
  \href {http://adsabs.harvard.edu/abs/2015MNRAS.447..140V} {447, 140}

\makeatother
\end{thebibliography}



\appendix 

\section{Convergence of Isolated Dwarf Galaxies} 

\begin{figure*} 
\centering 
\includegraphics[scale=0.42]{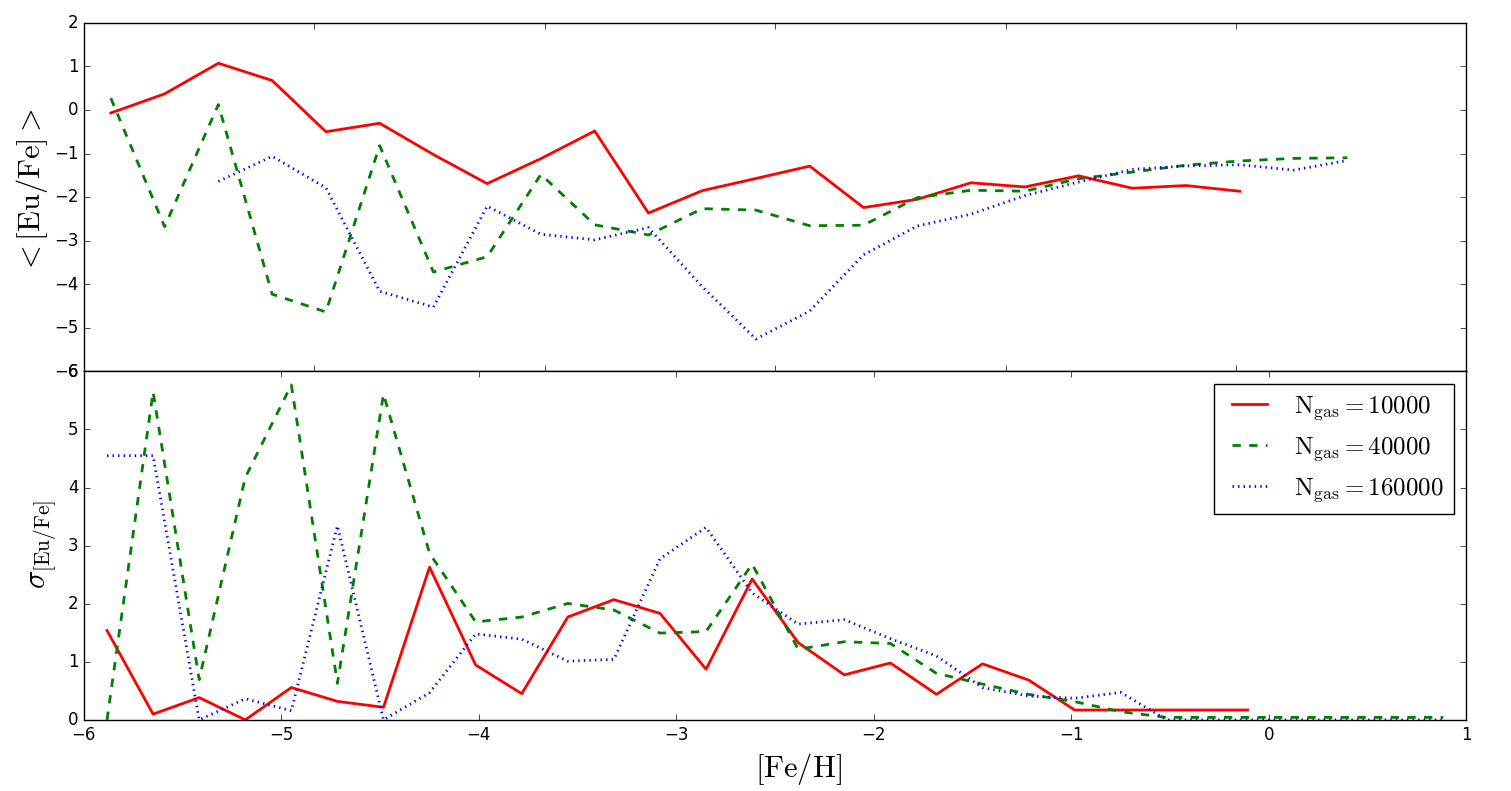} 
\caption{The upper panel shows the mean [Eu/Fe] as a function of [Fe/H] for three different resolutions: low resolution (10000 particles), medium resolution (40000) and high resolution (160000). The lower panel shows the corresponding standard deviation.} 
\label{DiscComp1} 
\end{figure*}

We show here the convergence of our simulations for our isolated dwarf disc galaxies. Figure \ref{DiscComp1} shows the mean <[Eu/Fe]> (upper panel) and standard deviation $\mathrm{\sigma_{[Eu/Fe]}}$ (lower panel) for three resolutions: low, medium and high, corresponding to $\mathrm{N_{gas}}$ = 10000, 40000 and 160000 gas particles in the initial conditions respectively, with the high resolution being what we use in Section 3. All three resolutions are very similar at [Fe/H] $\sim$ $-2$ and the medium and high resolutions give a very good match at higher metallicity. In the low resolution run there are no very metal rich stars because most of the gas particles have been ejected from the system due to small particle numbers and strong feedback. We note that this effect is not seen in larger galaxies (such as Milky Way type galaxies) due to the substantially greater number of particles and weaker feedback. At [Fe/H] $\la$ $-3$ there is a difference in <[Eu/Fe]> and $\mathrm{\sigma_{[Eu/Fe]}}$ depending on the resolution. This is likely caused by the small number of star particles at low metallicity. We note that these resolution dependencies may be different for the Milky Way simulations shown in Section 4 since they have a lower resolution and different initial conditions. 

\section{[O/Fe] Bi-modality} 

\begin{figure*} 
\centering 
\includegraphics[scale=0.38]{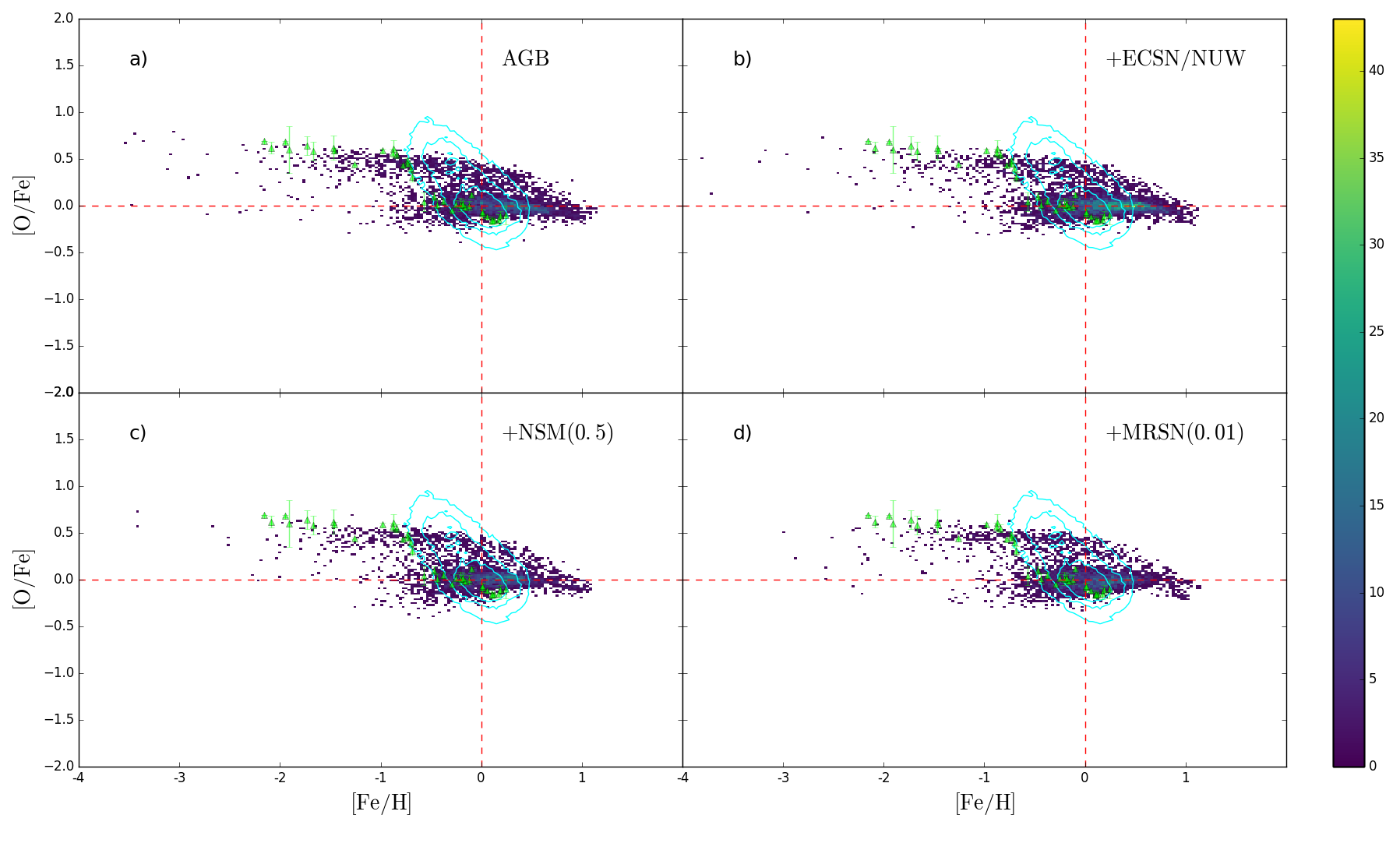} 
\caption{The same as Figure \ref{EUFEMW1} but with [O/Fe] plotted against [Fe/H].} 
\label{OFE1} 
\end{figure*}

The simultaneous agreement both with the observed [Eu/Fe] and [Eu/O] ratios means that we also have good agreement with the observed [O/Fe] ratios. Figure \ref{OFE1} shows the [O/Fe]--[Fe/H] relation for the models from Section 4. Our prediction is in excellent agreement with the recent non-local thermal equilibrium (NLTE) abundance observations from \citet{ZhaoEtAl16} (green triangles) and in very good agreement with the HERMES-GALAH survey (blue contours). The [O/Fe] distribution is similar across all these models and they present a bi-modality at [Fe/H] $\sim$ $-1$ as in the chemodynamical simulation by \citet[][Figure 10]{Kobayashi11b} and similar to the Ref-L100N1504 simulation galaxies from \citet{Mackereth18}. The details including the macroscopic mixing discussed in \citet{Mackereth18} will be included in future work (Haynes \& Kobayashi, in prep.). 

\section{Kernel-based mixing} 

\begin{figure} 
\centering 
\includegraphics[scale=0.40]{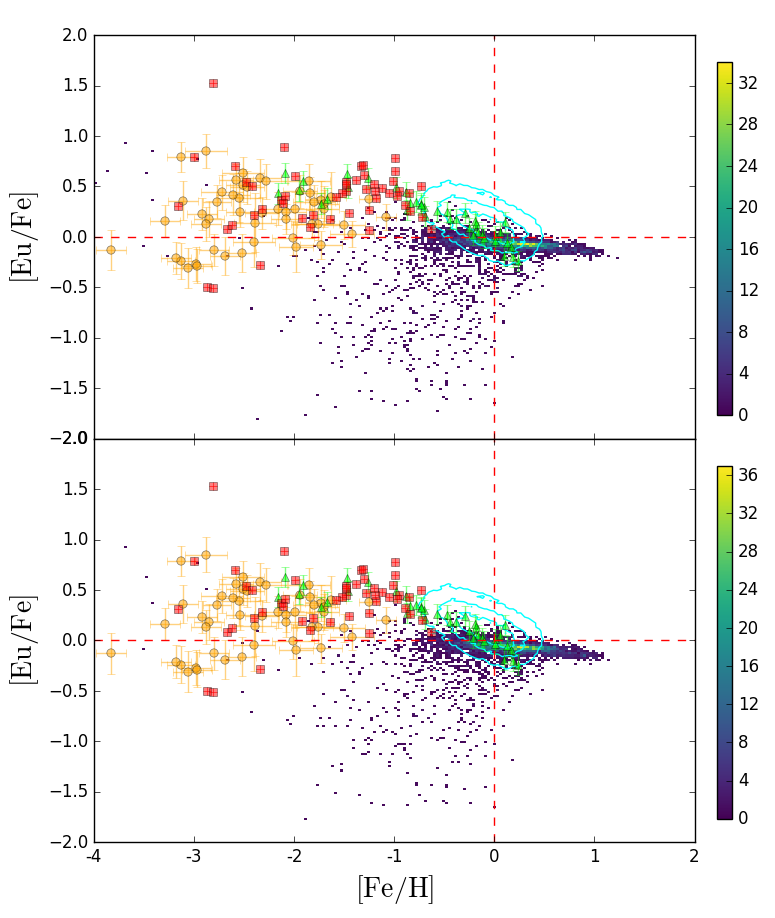} 
\caption{The same as Figure 7c (upper panel) but compared to the simulation with kernel smoothing (lower panel).} 
\label{Kernel_smoothing} 
\end{figure} 

Although we include no explicit sub-grid diffusion (Section 2.1), it is possible to examine an approximation of the effects of diffusion by using the kernel weighting to average the metallicities of nearby particles (\citealt{Crain13}). \citet{Mackereth18} found no difference for their [$\mathrm{\alpha}$/Fe]--[Fe/H] relation with the addition of kernel smoothing. Figure \ref{Kernel_smoothing} shows the [Eu/Fe]--[Fe/H] relation in the NS merger simulation with kernel smoothing (smoothing length of 0.2 kpc, lower panel) and the unsmoothed simulation (upper panel, the same as Figure 7c). We see no significant changes to the [Eu/Fe]--[Fe/H] relation after the kernel smoothing. We also find that this kernel smoothing does not significantly alter the [O/Fe]--[Fe/H] relations, which is in good agreement with observation, and maintains the bi-modality (see Appendix B).  It should be noted however that using an increased smoothing length ($\sim$ 1 kpc) can decrease the number of low [Fe/H] stars substantially and the [O/Fe]--[Fe/H] relations become inconsistent with observations.


\bsp  
\label{lastpage}
\end{document}